\documentclass[sigplan,nonacm,twocolumn]{acmart}

\usepackage{tikz}
 \usetikzlibrary{patterns}
\usepackage{threeparttable}
\usepackage[utf8]{inputenc}
\usepackage{graphicx}
\usepackage{subcaption}
\usepackage{fancyvrb}
\usepackage{wasysym}
\usepackage{tabularx}
\usepackage{multirow}
\usepackage{makecell}
\usepackage{array, multirow}
\usepackage{arydshln}
\newcommand{\ignore}[1]{}
\usepackage{balance}
\usepackage{xcolor}
\usepackage[ruled,vlined]{algorithm2e}
\usepackage{booktabs}
\usepackage{cleveref}
\usepackage{float}
\usepackage{flushend}
\usepackage{pgfplots,pgfplotstable}
\usepackage{subcaption}
\usepackage{amsmath}
\usepackage{listings}
\usepackage{xcolor}
\usepackage{hyperref}

\lstdefinestyle{example}{
    backgroundcolor=\color{gray!10},
    basicstyle=\ttfamily\small,
    frame=single,
    framesep=5pt,
    breaklines=true,
    xleftmargin=10pt,
    xrightmargin=10pt,
    numbers=none,
    showstringspaces=false,
    columns=fullflexible,
    keepspaces=true
}

\newcommand{\resizerop}{\mathrel{\scriptstyle \downarrow \hspace{-0.2em} \scriptstyle \uparrow}}

\newcommand*\halfcircbottom[1][1ex]{%
  \begin{tikzpicture}[baseline={([yshift=-0.2ex]current bounding box.center)}]
  \draw[fill] (0,0) -- (180:#1) arc (180:360:#1) -- cycle;
  \draw (0,0) circle (#1);
  \end{tikzpicture}}

\newcommand*\emptycirc[1][1ex]{%
  \tikz[baseline={([yshift=-0.2ex]current bounding box.center)}] \draw (0,0) circle (#1);}

\newcommand*\fullcirc[1][1ex]{%
  \tikz[baseline={([yshift=-0.2ex]current bounding box.center)}] \filldraw (0,0) circle (#1);}

\usepackage{enumitem}
\setlist{noitemsep}

\usepackage{pifont}

\newcolumntype{P}[1]{>{\centering\arraybackslash}p{#1}}

\usepackage{array}
\newcolumntype{H}{@{}>{\setbox0=\hbox\bgroup}c<{\egroup}}

\usepackage{tikz}
\usepackage{pgfplots,pgfplotstable}
\usepackage{subcaption}

\newcommand{\mywrap}[0]{{Reflex}}
\newcommand{\resizer}[0]{{Resizer}}

\renewcommand\footnotetextcopyrightpermission[1]{} 
\begin{document}
\title{\textsc{\mywrap}: Faster Secure Collaborative Analytics via Controlled Intermediate Result Size Disclosure}

\author{Long Gu}
\affiliation{\small{Systems Group, TU Darmstadt, Germany}}

\author{Shaza Zeitouni}
\affiliation{\small{Systems Group, TU Darmstadt, Germany}}

\author{Carsten Binnig}
\affiliation{\small{Systems Group, TU Darmstadt, Germany}}

\author{Zsolt István}
\affiliation{\small{Systems Group, TU Darmstadt, Germany}}
\begin{abstract}
Secure Multi-Party Computation (MPC) enables collaborative analytics without exposing private data. However, OLAP queries under MPC remain prohibitively slow due to oblivious execution and padding of intermediate results with filler tuples. 
We present \textsc{\mywrap}, the first framework that enables \emph{configurable trimming of intermediate results across different query operators}—joins, selections, and aggregations—within full query plans. At its core is the \emph{\resizer} operator, which can be inserted between any oblivious operators to selectively remove filler tuples under MPC using user-defined probabilistic strategies. 
To make privacy trade-offs interpretable, we introduce a new metric that quantifies the number of observations an attacker would need to infer the true intermediate result sizes. \textsc{\mywrap} thus makes the performance–privacy space of secure analytics \emph{navigable}, allowing users to balance efficiency and protection. Experiments show substantial runtime reductions while maintaining quantifiable privacy guarantees.
\end{abstract}

\maketitle

\section{Introduction}

\begin{figure}
  \centering
  \small
    \begin{subfigure}{0.45\linewidth}
        \centering
        \includegraphics[width=\linewidth]{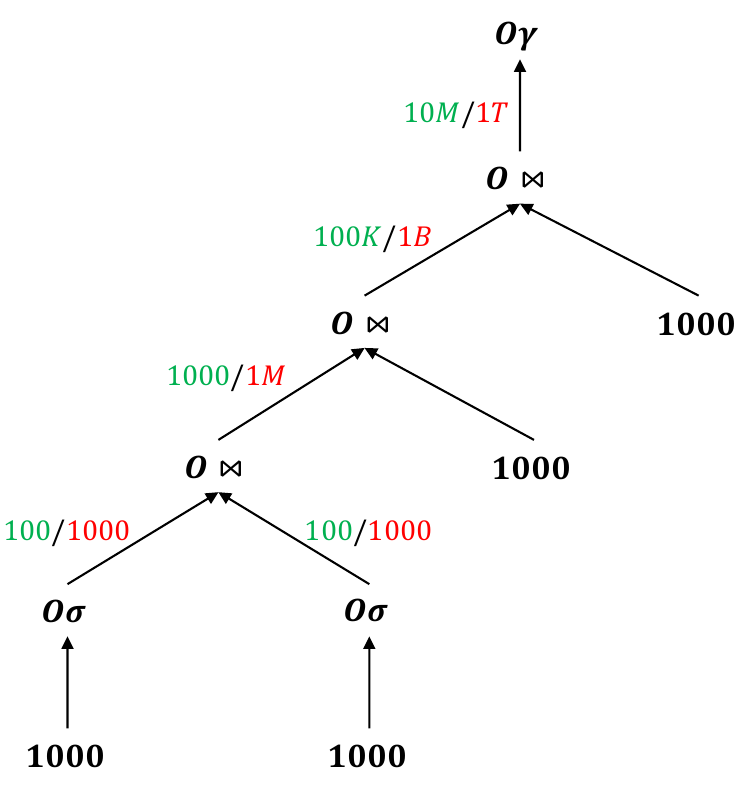}
        \caption{3-join Query Plan, with (red) and without (green) oblivious output sizes}
        \label{fig:3join_qp}
    \end{subfigure}%
    \hfill
      \begin{subfigure}{0.52\linewidth}
      \centering
      \begin{tikzpicture}
      \begin{axis}[
          clip = false,
          xlabel={Amount of Filler Tuples ($\% N$)},
          ylabel={Total Tuples to Process},
          y label style={at={(axis description cs:0.15,0.5)}, anchor=south},
          width=\linewidth,
          height=5cm,
          grid=both,
          xmin=0, xmax=10,
          ymin=-2e11,
          xtick={0,...,10},
          xticklabels={0\%,10\%,20\%,30\%,40\%,50\%,60\%,70\%,80\%,90\%},
          xticklabel style={rotate=90, anchor=base, xshift = -2mm}, 
          x tick label style={font=\scriptsize},
      ]
        \addplot[blue, thick, mark=*] coordinates {
            (0, 101010400) (1, 1608040800) (2, 8127091200) (3, 25664161600)
            (4, 62625252000) (5, 129816362400) (6, 240443492800)
            (7, 410112643200) (8, 656829813600) (9, 1001001004000)
        };
        \addplot[only marks, mark=*, mark options={fill=green!60!black, draw=green!60!black}]
        coordinates {(0,101010400)};
        \addplot[only marks, mark=*, mark options={fill=red, draw=red}]
        coordinates {(9, 1001001004000)};
        
        \draw[dotted, thick] (axis cs:0,0) -- (axis cs:9,0);
        \draw[dotted, thick] (axis cs:0,1e12) -- (axis cs:9, 1e12);
    
        \node[font=\scriptsize, anchor=south east, text=red] at (axis cs:8.8,0.98e12) {Fully Oblivious};
    
        \draw[<->, blue, thin, bend right=15] (axis cs:3,2.3e11) to (axis cs:7,7.8e11);
        \node[blue, font=\scriptsize, anchor=south east, align=right, text width=2.8cm] at (axis cs:5.8,6e11) {\textsc{Reflex}: navigate\\ the space};

        \draw[->,dashed, black] (rel axis cs:0.5, 1.05) -- (axis cs: 8.9,10.3e11);

        \draw[->, dashed, black] (rel axis cs:0.5, 1.05) -- (axis cs: 0,0.4e11);

        \node[above, right=0.5pt, font=\tiny, black] at (axis cs:1.7, 1.24e12) {Existing point solutions~\cite{scql_fangsecretflow,secrecy_liagouris2023secrecy,smcql}};
        
        \draw[->, orange, thick] 
           (axis cs:7.7,1.3e11) -- (axis cs:9.5,6.8e11) 
           node[midway, above, sloped, orange, font=\scriptsize] {Exponential blow-up};
        \node[font=\scriptsize, text=green!60!black, anchor=west, align=right] at (axis cs:0.1,-1e11) {Revealed};

      \end{axis}
      \end{tikzpicture}
      \caption{3-join's total tuples processed under varying strategies.}
      \label{fig:3join_tc}
    \end{subfigure}

  \caption{Motivating 3-join example: all operator selectivities are fixed at $10\%$, and the total number of tuples to process (sum of all intermediate result sizes) is a function of the amount of oblivious filler tuples included in each intermediate result. The trend is exponential when going from \textit{no fillers} (green) to \textit{fully oblivious intermediate sizes} (+$90\% N$ filler tuples at each operator, red)-- this explains the severe performance penalty under fully-oblivious MPC.}
\end{figure}

The analysis of internal data is a critical component of strategic decision-making in large organizations. Since most of these organizations operate globally, transferring and processing data across countries and jurisdictions without strong cryptographic protection introduces risks and may even be restricted by regulations such as the GDPR~\cite{GDPR2016general}. As a result, there is a growing effort~\cite{smcql,conclave_volgushev2019conclave,secrecy_liagouris2023secrecy,senate_yanai2021senate} to combine analytical data processing with Secure Multi-Party Computation (MPC) in order to provide stronger security and privacy guarantees.
MPC allows several parties with private data, e.g., company branches, to jointly compute a function, e.g., a database query, without any party learning more in the process than the final result. 


However, performing computations under MPC is several orders of magnitude more expensive than in plaintext~\cite{smcql,he2021practical_security_privacy_db,roy2020crypt_crypt_system_dp,secrecy_liagouris2023secrecy}.
In addition to the per-operation computation overhead, there is a compounding factor for OLAP-like workloads: algorithms in MPC must execute \textit{obliviously}, i.e., without revealing any information about the underlying data. For queries, this means that intermediate results passed between operators of a query plan must be padded with filler tuples to the maximum possible size, preventing attackers from inferring information based on the selectivity of the operators. For example, selection operators must return an output of the same size as the input, and joins must produce an output with the size of the Cartesian product of both tables. Overall, for query plans with multiple operators, this can lead to result size exploitation as shown in Figure~\ref{fig:3join_qp}. 

In this paper, we present \textsc{\mywrap}, a novel approach that accelerates private query processing for OLAP workloads.
The key idea behind \textsc{\mywrap} is to make the \emph{performance--privacy trade-off} in query execution both visible and controllable. It is the first method that allows even non-experts to \emph{quantitatively} explore how much performance can be gained for a given level of privacy protection.
At the heart of \textsc{\mywrap} is a new query operator called the \emph{\resizer}. This operator can be placed between any pair of oblivious operators in a query plan. It reduces the size of intermediate results by \emph{randomly trimming filler tuples}---dummy records added for privacy. The trimming amount is determined through \emph{secure sampling under MPC}, guided by user-defined probability distributions that capture different performance--privacy preferences.
To measure the amount of privacy each distribution provides, we introduce a new metric that quantifies how many repeated observations of the trimmed intermediate result sizes an attacker would need to accurately infer the true intermediate result size.

The concept of trimming intermediate results has been explored in several recent works. However, existing approaches have primarily focused on the \emph{join} operator~\cite{shrinkwrap_bater2018shrinkwrap,chang2022towards_oblivious_join} and do not integrate all mechanisms into a single unified framework to make privacy quantifiable at the query plan level. To the best of our knowledge, there is currently no \emph{general mechanism} that applies uniformly across different query operators (e.g., join, selection, aggregation), integrates seamlessly into query plans without modifying existing oblivious operators, and enables fine-grained control over the performance--privacy trade-off at the operator level.
Moreover, prior works typically employ fixed trimming strategies tailored to specific privacy assumptions, such as differentially private resizing~\cite{shrinkwrap_bater2018shrinkwrap}. We argue that a practical analytics framework should instead provide both \emph{flexibility} and a \emph{unified means} to quantify the privacy guarantees of user-defined trimming strategies for intermediate result sizes.

To achieve the goals of \textsc{\mywrap}, we address two key challenges in this paper. 
First, it requires a \emph{generic, configurable, and efficient} resizing mechanism that can be integrated into diverse query plans. 
The \resizer~must yield tangible performance gains by reducing upstream computation, without introducing overheads that offset these benefits. 
Achieving both configurability and efficiency across heterogeneous operators demands careful algorithmic and system-level co-design. 
Second, \textsc{\mywrap} must address a \emph{usability challenge}: how to provide non-security experts with an intuitive means to quantify and navigate the performance--privacy trade-off introduced by trimming filler tuples. 
To this end, we develop a practical metric that abstracts away cryptographic complexity while capturing how resizing choices affect both system efficiency and potential leakage of intermediate result sizes. 

Addressing these challenges within a unified framework enables a principled exploration of the performance--privacy space, moving beyond static, one-size-fits-all oblivious execution. 
As illustrated later in \Cref{fig:privacy_vs_runtime_points},
different placements and configurations of the \resizer{}~yield predictable trade-offs between runtime and privacy. 
Our experiments show that trimming can reduce query runtimes by up to an order of magnitude, while an attacker would still require multiple rounds of observation to infer the true intermediate sizes.
Ultimately, this line of work can pave the way for \emph{MPC-aware query optimizers} that jointly explore the space of oblivious query plans and resizing strategies to find configurations that maximize performance while meeting certain user-defined security guarantees. Building such an optimizer, however, is beyond the scope of this paper. 
To summarize, our contributions are as follows:
\begin{itemize}
\item
\textbf{A first framework for privacy–performance trade-offs}. We present \textsc{\mywrap}, a framework that provides fine-grained, user-controlled trade-offs between performance and privacy for Secure collaborative analytics  (SCA). The central primitive is the \resizer, a lightweight operator that can be inserted after any oblivious operator to selectively remove filler tuples according to user-specified trimming strategies while preserving obliviousness. \resizer’s MPC-friendly design increases parallelism and reduces communication rounds, enabling efficient and flexible reduction of intermediate result sizes. We implement a prototype and release the code as open-source; see \Cref{sec:artifacts}.

\item
\textbf{A statistically grounded metric.} To enable principled comparison between different user-defined trimming approaches, we propose a metric grounded in statistical methods that quantifies how many observations of trimmed intermediate result sizes an attacker would need to recover the true intermediate result size with high probability. The metric supports direct, interpretable comparisons between different trimming strategies and can be used to place points in the privacy–performance design space.

\item
\textbf{A comprehensive evaluation}. We thoroughly evaluate \textsc{\mywrap}~through micro-benchmarks, reimplementation \& comparison with related methods~\cite{shrinkwrap_bater2018shrinkwrap,scql_fangsecretflow}, as well as full query experiments using analytical queries from prior work. Our study characterizes runtime across multiple points in the trade-off space, and demonstrates the practical gains and limits of different trimming strategies under both performance and privacy metrics.
\end{itemize}

\section{Background and Related Work}

\subsection{Secure Collaborative Analytics}
\label{subsec:system_model}

Secure collaborative analytics  (SCA)~\cite{secrecy_liagouris2023secrecy} is a set of distributed data statistics protocols designed to protect private datasets provided by \textit{Data Owners}.
Let \(\mathcal{DO} = \{DO_1, \ldots, DO_\ell\}\) be a set of $\ell$ \textit{Data Owners}, where each $DO_i$ holds a private database $D_i$. 
 Secure Collaborative Analytics protocols enable joint data analysis on the union of all private datasets \(\mathcal{D} = \{D_1, \ldots, D_\ell\}\) while safeguarding against privacy violations.
Let \(\mathcal{P} = \{P_1, \ldots, P_m\}\) denote a set of $m$ computing nodes or parties that are responsible for executing secure collaborative analytics protocols on $\mathcal{D}$. The \textit{Data Analyst} sends the query $Q$ to be executed over the union of all private datasets $\mathcal{D}$ to $\mathcal{P}$ and collects the final result $R$. Overall, SCA is a growing field with numerous real-world applications, including market analysis, autonomous driving, agriculture, and the Internet of Things. These applications benefit from efficient query execution among multiple data owners, which also preserves data privacy~\cite{Bosch_use_case}.

\subsection{Related Work using MPC} 
To the best of our knowledge, there is no related work that presents a flexible framework for balancing the trade-off between revealing information about intermediate result sizes in query execution in a semi-honest setting. The related work in SCA using MPC can be grouped into one of three groups: first, those that aim to optimize the underlying oblivious operators given metadata about the query, second, those that offer point solutions in terms of relaxing security guarantees, and third, those that aim to protect against a much stronger, malicious attacker. 

Related work, such as SMCQL~\cite{smcql}, Secrecy~\cite{secrecy_liagouris2023secrecy}, Senate~\cite{senate_yanai2021senate}, and Alchemy~\cite{sohn2025_alchemy}, aims to increase performance by leveraging relations' metadata and redesigning actual computation (hence also query execution) under MPC or applying algorithmic changes to the MPC protocols without loosening security/privacy guarantees. 
One of the most recent works in this direction is Alchemy \cite{sohn2025_alchemy}.  Alchemy optimizes queries through rewrite rules, cardinality bounds, bushy plan generation, and a fine-grained cost model, minimizing circuit complexity while maintaining security guarantees. Such optimizations are orthogonal to our work and can be adopted in \textsc{\mywrap}~to further improve performance.


Other related work, such as Secretflow~\cite{scql_fangsecretflow}, Shrinkwrap~\cite{shrinkwrap_bater2018shrinkwrap}, and SAQE~\cite{saqe_bater2020SAQE}, improves performance by relaxing the privacy guarantees, such as information about the intermediate result size of operators~\cite{shrinkwrap_bater2018shrinkwrap,scql_fangsecretflow}, or the tuples that form the intermediate result~\cite{scql_fangsecretflow}. Some works even allow for relaxing correctness of query results~\cite{shrinkwrap_bater2018shrinkwrap,saqe_bater2020SAQE}.
Closest to the approach of \textsc{\mywrap} is Shrinkwrap \cite{shrinkwrap_bater2018shrinkwrap}, which introduces a point solution that reduces fully-oblivious intermediate result sizes to differentially private sizes. This is achieved with a ``sort\&cut'' approach, which couples trimming with a potential reduction in query accuracy (no guarantee of correctness). In the context of analytics within an enterprise, this is not desirable -- in \textsc{\mywrap}, queries always produce the same results, regardless how many filler tuples have been removed from the intermediate result. 

SAQE \cite{saqe_bater2020SAQE} employs a private sampling algorithm that trades correctness for performance, providing a scalable and fast approximate query processing platform. 

Finally, there is also related work that focuses not on performance improvements but on protecting against a stronger attacker in the malicious security model, e.g., Senate \cite{senate_yanai2021senate} and Scape \cite{scape_han2022scape}. 
Since protocols that are secure against a malicious attacker are yet again orders of magnitude slower than those for semi-honest systems \cite{Lindell19,dimitriou2014multi}, and since these systems introduce restrictive assumptions about how data can be processed, they are not a perfect match for the enterprise-level analytics we are focusing on.

\subsection{Related Work using TEEs} In case the security model permits a trusted third party, Trusted Execution Environments (TEEs), such as Intel SGX~\cite{costan16sgx,mckeen13innovative} or AMD SEV~\cite{kaplan2016amd}, can be used to perform fast analytics on shared data. Sensitive data is encrypted and only decrypted and processed within the TEE. There is rich related work, such as EnclaveDB~\cite{enclavedb_sgx_priebe2018enclavedb}, ObliDB~\cite{sgx_oblidb_eskandarian2017oblidb}, and benchmarking papers~\cite{lutsch_sgxbenchmark_edbt25}, showing that TEEs can be used for data analytics tasks with minimal performance overhead compared to plaintext execution. Compared to MPC, TEE-based methods are orders of magnitude faster. However, their adoption is restricted to the use of specialized hardware and vulnerabilities such as side-channel attacks \cite{sgx_sidechannel_lee2017inferring,sgx_sidechannel_brasser2017software,batteringramsp26_sgx_attack_battery_ram, wiretap_sgx_attack}.

\begin{figure*}[!ht]
    \centering
    \begin{subfigure}[b]{0.3\linewidth}
        \centering
        \includegraphics[width=0.7\linewidth]{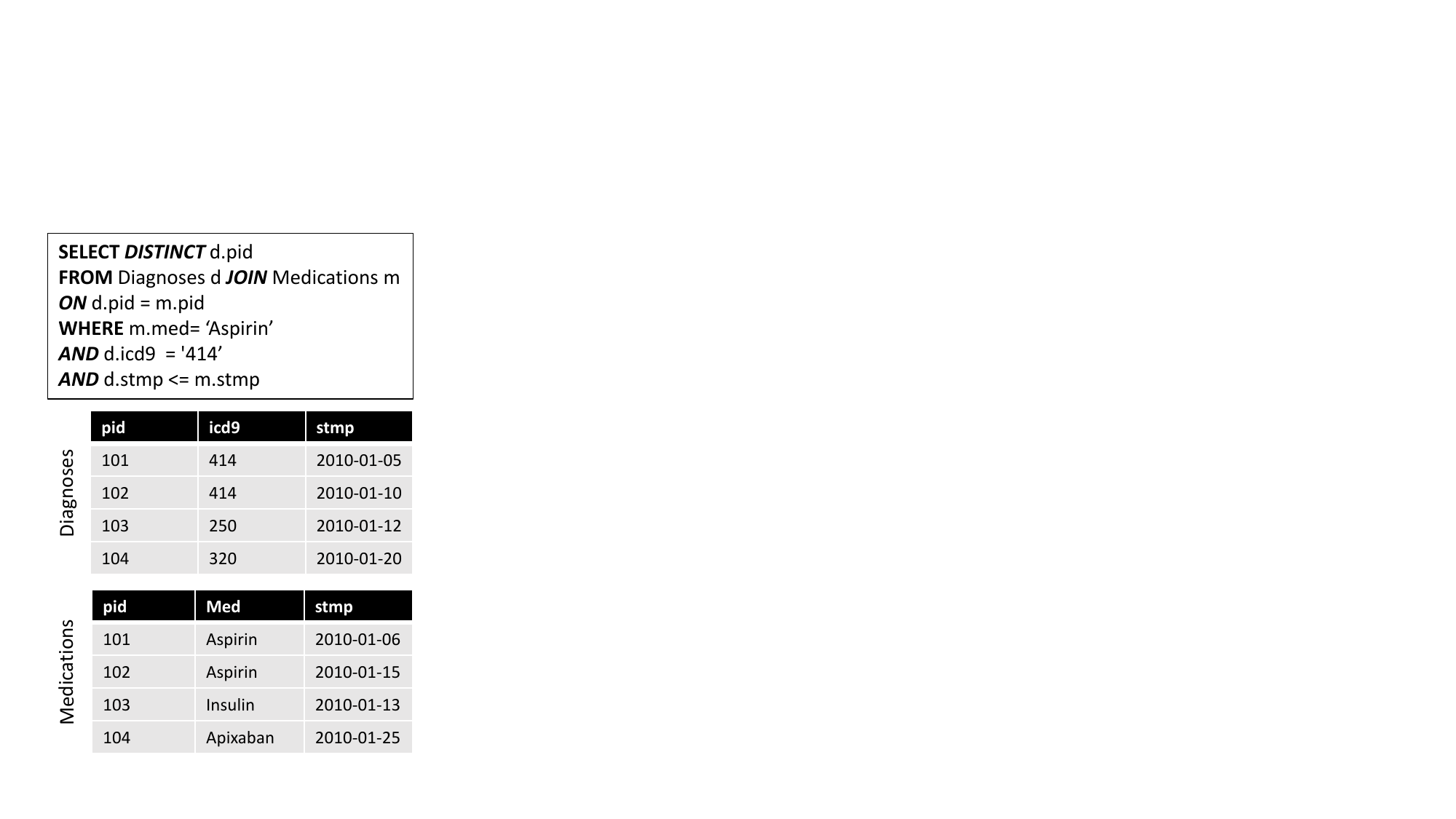}
        \caption{Example Query}
        \label{fig:sub1}
    \end{subfigure}
    \begin{subfigure}[b]{0.3\linewidth}
        \centering
        \includegraphics[width=0.8\linewidth]{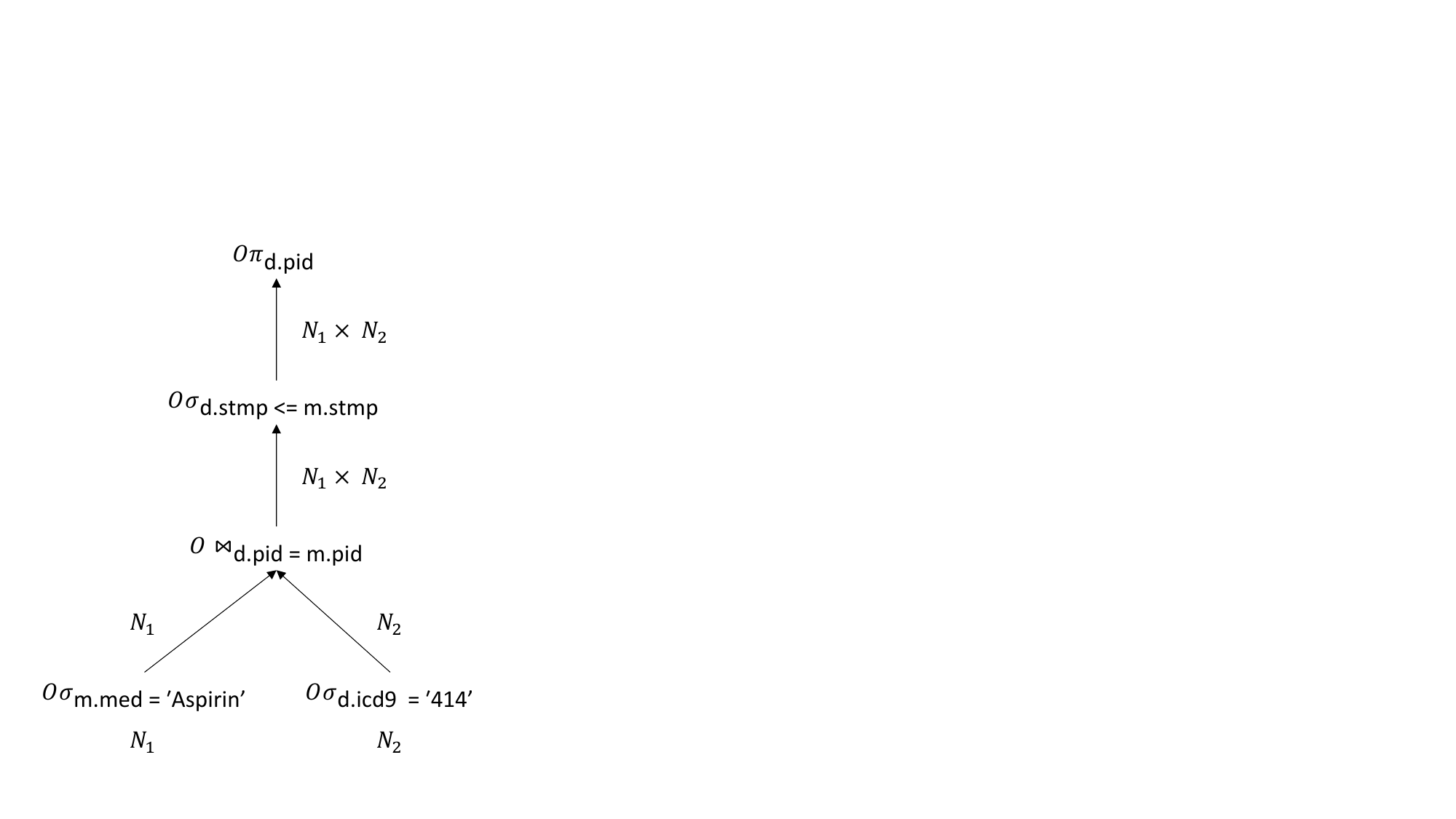}
        \caption{Fully Oblivious Query Plan}
        \label{fig:sub2}
    \end{subfigure}
    \begin{subfigure}[b]{0.3\linewidth}
        \centering
        \includegraphics[width=0.8\linewidth]{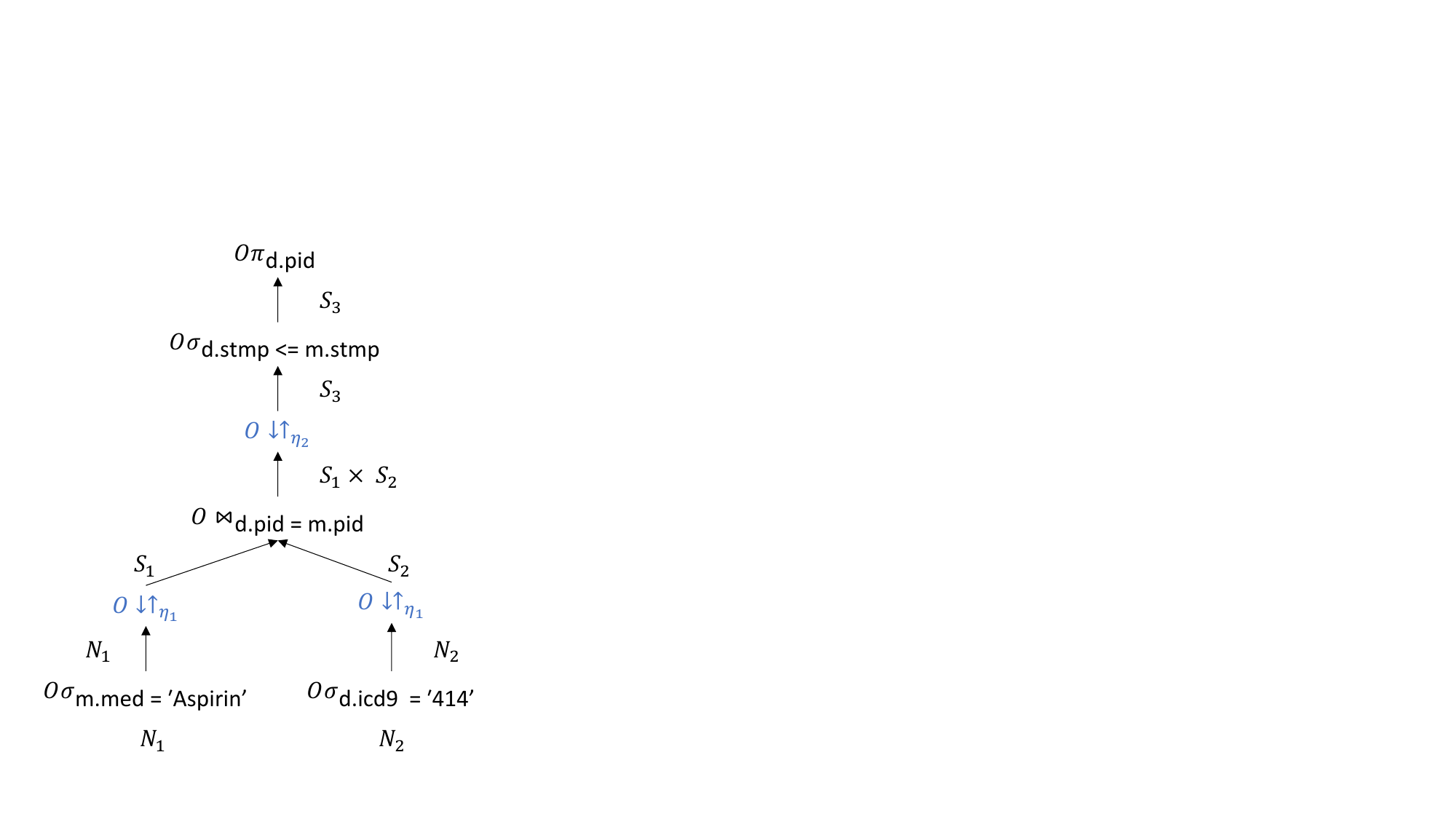}
        \caption{Oblivious Query Plan with \resizer~($\resizerop$)}
        \label{fig:sub3}
    \end{subfigure}

    \caption{A Query Example and its Plans. $O$ indicates oblivious operators, $N_i$ is input/output sizes, and $S_i$ refers to re-sized output after $\resizerop$ operator with ($S_i < N_i$).}
    \label{fig:main}
\end{figure*}

\subsection{Related Work using FHE}

There are also related works based on Fully Homomorphic Encryption (FHE), such as HE3DB~\cite{2023HE3db}. It allows multiple data owners to encrypt their private input and conduct collaborative analytics in the cloud server without exposing intermediate results. 
HE3DB~\cite{2023HE3db} utilizes the latest advancements in FHE~\cite{FHE_guimaraes2021revisiting} and allows for server-side elastic analytical processing of requested FHE ciphertexts, including private decision tree assessment, unlike previous encrypted DBMS that only provide aggregated information retrieval.
While FHE provides stronger security guarantees than MPC, sparking growing interest, its significant performance overhead currently renders it impractical for complex operations, such as secure sorting or joins, which are at the core of SCA.

\section{\textsc{\mywrap}: Assumptions and Threat Model}
\label{sec:assumption}
In SCA, the goal is to protect the privacy of input datasets and computation. Ideally, no party should learn more about the data than what can be inferred from the final query result. 

In the deployment model as described in \Cref{subsec:system_model}, each data owner, $DO_i$, distributes secret shares of their dataset, $D_i$, to the computing nodes. A query $Q$ is translated into a set of MPC protocols $\pi$, which are executed by all computing nodes $\mathcal{P}$ on the secret shares of the union of all datasets $\mathcal{D}$.
It is a common assumption that under MPC, all parties, i.e., data owners, computing nodes, and the data analyst, are assumed to know the database schema and the queries to be executed.
During the execution of $\pi$ on input data shares of $\mathcal{D}$, each computing node $P_i$ has a view that includes the following:
\begin{itemize}
\item $P_i$'s secret share of $\mathcal{D}$ and any locally generated random values required by the protocol,
\item messages received during protocol execution,
\item $P_i$'s secret shares of intermediate results,
\item \textit{trimmed} sizes of intermediate results. These are some values between true intermediate result size and fully-oblivious result size, and
\item $P_i$'s secret shares of the final response $R$.
\end{itemize}

We assume a \textit{semi-honest} trust model, where computing nodes follow the protocol honestly but may attempt to infer information from their local view about the input datasets, intermediate results, or the sizes of intermediate results of query operators. 
This trust model is suitable for protecting against accidental data leaks, such as logs exposed to system administrators or lost disks. and it aligns well with the context of a large corporation (similar to the use cases presented in~\cite{scql_fangsecretflow,Bosch_use_case}), where data owners and computing nodes belong to branches or departments within the same large corporation. The goal of using secure MPC in this context is to protect the privacy of computation and data in motion, to contribute to data protection by design~\cite{dpbd}.

On top of the semi-honest trust model, we adopt the following adversary model, where: \emph{(i)} at most one computing node may be corrupted,
and \emph{(ii)} parties do not collude.
Therefore, it is reasonable to assume an attacker can \textit{only} observe, without invasive interference, the trimmed intermediate result sizes of all operators due to the insertion of \resizer~operators across queries executed in the system and use these observations of trimmed intermediate result sizes to learn the true output size of an operator.

As a way to quantify the information leakage as a result of trimming filler tuples from intermediate results, we introduce the \textit{Rounds to Recover} metric (detailed in \Cref{sec:security}) that determines the number of equivalent repetitions of an operator required to infer its true result size with high probability. 
An equivalent repetition refers to any execution, possibly across different queries, where the operator processes the same input and produces the same intermediate result. This metric provides non-security experts with an intuitive way to compare the trade-off between performance and privacy of different strategies that \textsc{\mywrap}~allows them to define. Note that in this work, we do not study the effect of revealing trimmed intermediate result sizes on the privacy of datasets. Therefore, in the remainder of the paper, by `privacy' we mean the privacy of intermediate result sizes.


\section{\textsc{\mywrap}: Design and Implementation}
\label{sec:mywrap}

In the following, we provide an overview of \textsc{\mywrap} before delving into the details of its core mechanisms and the metric used to quantify the security gains.

\subsection{Overview of \textsc{\mywrap} }
\textsc{\mywrap}~is a framework that provides fine-grained, user-controlled trade-offs between performance and privacy for SCA. This is achieved with the help of \resizer~operators that can be inserted after any oblivious operator to remove filler tuples according to a user-specified trimming strategy. \resizer~has been implemented in a way that can take advantage of communication batching optimizations in MPC, with a runtime comparable to oblivious filter, join, and group by operators. See  Section~\ref{sec:resizer} for more details.

One of the major novelties in \textsc{\mywrap}~is the flexibility of deciding how many filler tuples are kept, without having to change the underlying \resizer~implementation. \Cref{fig:main} illustrates a simple query where \resizer~operators are inserted to reduce the output size of the filters following a scan and the output size of a join. Flexibility in configuration and placement enables \textsc{\mywrap}~to achieve significant performance improvements. 

This flexibility of placement is achieved by plugging in a user-defined strategy into each \resizer instance. The strategy is, at its core, a pre-defined probability distribution for determining how many filler tuples to keep in the intermediate result passed to the next operator. Note that the distribution can be re-configured as often as needed to fulfill performance or privacy requirements. In~\Cref{subsec:distr}, we discuss the properties a distribution should have to ensure some protection of the intermediate result size, and present concrete examples of distributions and how they are sampled. 

Without a straightforward way to compare user-defined distributions, it is not really possible to argue for specific ones. To this end, in Section~\ref{sec:security} we present the metric we propose to compare distributions in terms of how much information they leak.

\textbf{Our Implementation:} For our implementation, we utilize MP-SPDZ~\cite{keller2020mp-spdz}, a powerful framework for secure MPC, which compiles applications written in a Python-like code into MPC protocols. In addition to offering state-of-the-art performance, MP-SPDZ enables the selection of different threat models and underlying cryptographic primitives without requiring changes to the  code. 

To achieve efficient and secure computation, we choose Replicated Secret Sharing (RSS)~\cite{RSS_replicated_secret_sharing_araki2016high} as the underlying secret sharing scheme. RSS is highly efficient because it requires only a single round of communication for basic arithmetic operations, offering low latency compared to other protocols. Its integration within MP-SPDZ allows for seamless deployment and execution.
MP-SPDZ employs a mixed domain execution model. This allows the system to seamlessly convert secret shares between the arithmetic and binary (Boolean) domains using edaBits~\cite{escudero2020improved_edabits}. The internal compiler/optimizer automatically determines which domain is most efficient for executing a given function.

In addition to the \resizer~operator, our prototype  provides independent implementations of the following operators:
\begin{itemize}
    \item Scan: This operator reads private data into a secret shared matrix using for loop in MP-SPDZ~\cite{keller2020mp-spdz}. It assumes a known, fixed size for the dataset, allowing secure access and processing of each element. Note that a scan is always coupled with one of the other operators. 
    \item Filter: It applies a condition (multiple equality tests) to the entire table in a for loop, creating a new secret shared column marking matches. 
    \item Join: Implemented as a nested-loop join (NLJ), this operator uses two nested loops to compare every row from the first table (size n) with every row from the second table (size m). The column on which to evaluate the join predicate in each table is configurable. Joins produces an output of size $n * m$ and an extra secret column that marks true join result tuples. 
    \item Group By: Similarly to the related work~\cite{smcql,secrecy_liagouris2023secrecy}, we use a sorting-based method. We first order rows by the group key. Then, iterates row by row, comparing adjacent keys over secrets to identify group boundaries, marking them accordingly for subsequent aggregation operations. 
\end{itemize}

Queries are hand-assembled trees of SQL operators and instances of \resizer. Automated translation of plaintext SQL queries into MPC protocols remains future work.

\subsection{Resizer: A Helper Operator}
\label{sec:resizer}

\begin{figure}[t]
\centering
\includegraphics[width=0.75\columnwidth]{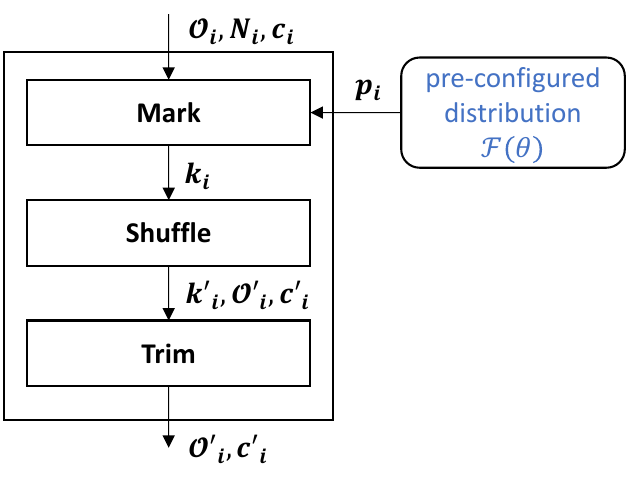}
\caption{Resizer operator. Inputs: oblivious output $\mathcal{O}_i$, oblivious output size $N_i$, true output column $c_i$ of operator $O_i$. Outputs: shuffled output $\mathcal{O}'_i$ and $c'_i$ indicating true tuples.}
\label{fig:resizer}
\end{figure}

A \resizer~operator can be inserted after any oblivious operator to reduce its output size by discarding filler tuples partially, without requiring query rewrites. It is built as a sequence of three processing steps: Mark, Shuffle and Trim (see Figure~\ref{fig:resizer}), which we detail in their own subsections.  The \resizer~runs under MPC and reveals no information about the contents of the output of the operator it runs after, apart from the final \textit{trimmed} output size $S_i= T_i + \eta_i$ where $T_i \leq S_i \leq N_i$ and $\eta_i$ is the amount of filler tuples retained by the \resizer. 


The amount of filler tuples to keep at each instance of \resizer~is decided at run-time (under MPC) using a pre-configured user-defined function (i.e., a distribution) that leverages both publicly-available information, e.g., input sizes and the operator trees, and secret information.



\begin{figure*}[t]
\centering
\includegraphics[width=\linewidth]{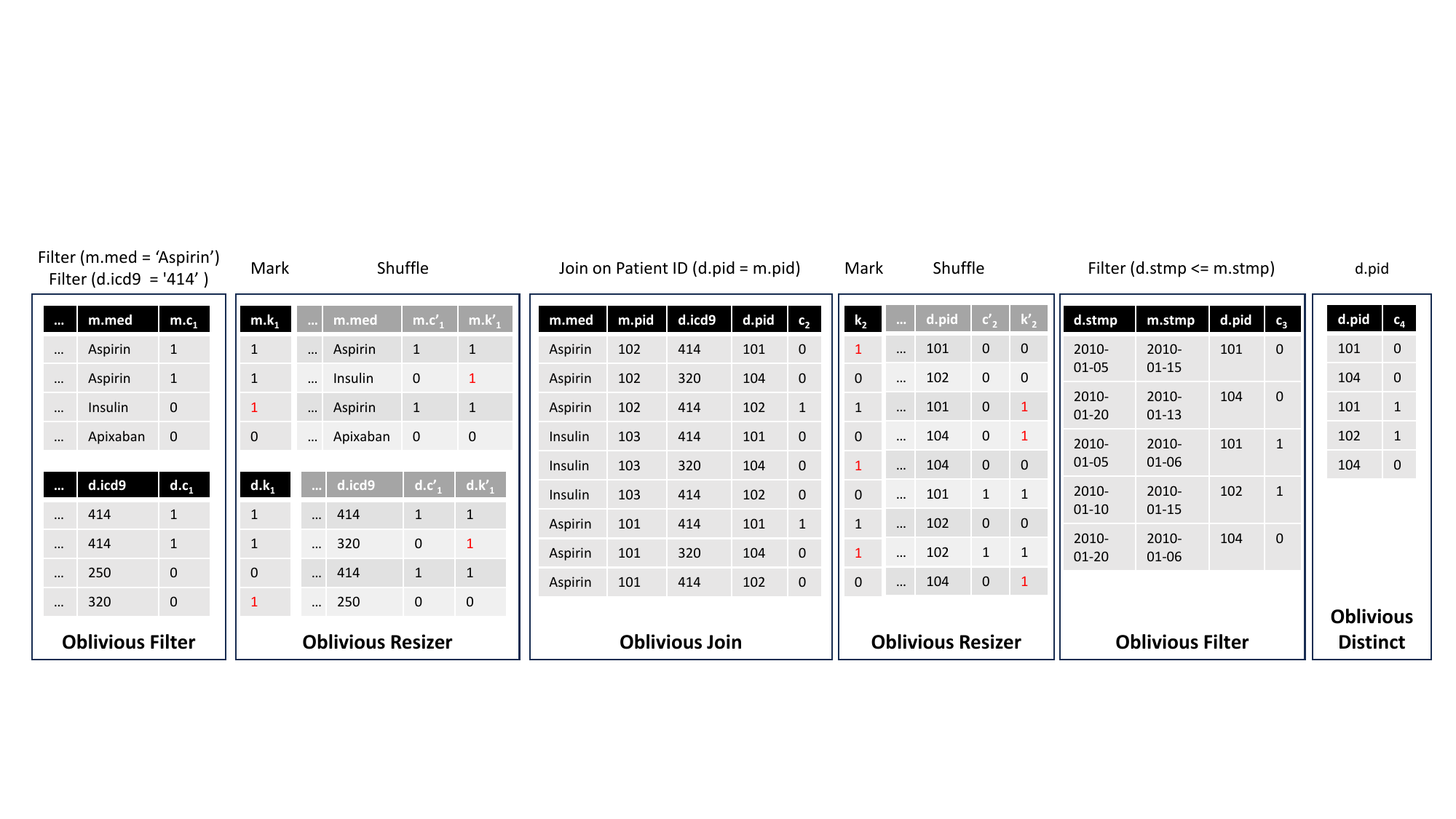}
\caption{Privacy-preserving query execution with \resizer~ operator. Retained filler tuples are shown in red.}
\label{fig:example}
\end{figure*}

\subsubsection{How we modify the output of oblivious operators?}

An oblivious operator $O_i$ in a query plan produces an oblivious output $\mathcal{O}_i = O_i(\mathcal{D})$ of size $N_i$ ($i$ indexes the operator in the query tree, e.g., \Cref{fig:sub3}). By construction, this output comprises $T_i$ tuples that are genuine output (called interchangeably ``true matches'') of the operator, while the remaining $N_i - T_i$ tuples are indistinguishable fillers that are retained as part of the output to satisfy obliviousness; we refer to these as \textit{filler tuples}.


The \resizer~(see \Cref{fig:resizer}) takes as input the oblivious output $\mathcal{O}_i =O_i(\mathcal{D})$, the oblivious output size $N_i = |\mathcal{O}_i|$, and the column $c_i$, which indicates whether a tuple is part of $O_i$'s true output. \resizer~takes also as an input a probability, $p_i$, sampled from a pre-configured distribution. $p_i$ will be used to determine which filler tuples to retain in the operator output. As output, \resizer~returns the shuffled oblivious output $\mathcal{O}'_i$ and the shuffled column $c'_i$, both trimmed to size $S_i = T_i+\eta_i$.   


Based on $p_i$, the \resizer~marks tuples to be retained from $\mathcal{O}_i$ -- note that true tuples are always marked. The marking is stored in column $k_i$, which is added to $\mathcal{O}_i$. 
In \Cref{fig:example}, filler tuples shown in red will be retained in the intermediate results: in columns $d.k_1$ and $m.k_1$ for the two \resizer~operators after the Filters, and in $k_2$ for the one after the join. 
The output $\mathcal{O}_i$, including all tuples, is then shuffled by $k_i$ to mitigate linkage attacks. Finally, tuples with $k'_i = 0$ are discarded, and $\mathcal{O}'_i$ contains only true and filler tuples of size $S_i = T_i + \eta_i$. As discussed before, $p_i$ is sampled obliviously and \resizer~operations are executed obliviously; therefore, $T_i$ and $p_i$ remain concealed, and only $S_i$ is revealed.  

In the following, we focus on how filler tuples are marked, describing how this step is optimized for efficient MPC execution. The choice of the distribution is deferred to \Cref{subsec:distr}.

\subsubsection{Marking Filler Tuples Before Trimming}
\label{subsec:noise_add}
The \textit{Mark} step (\Cref{fig:resizer}) takes as additional input the $p_i$ sampled from the user-defined distribution attached to the instance of the \resizer, the oblivious output size $N_i$ of the preceding SQL operator $O_i$, and the column $c_i$, which identifies true tuples. Mark step outputs a column $k_i$, indicating tuples to keep from trimming. 
In the following, we present our approach to efficiently carrying out the marking step.  


The marking of tuples to keep is done stochastically: a secret weighted coin (with $p_i$) is flipped $N_i$ times, with each flip determining whether a tuple is marked for keeping. This method offers a significant advantage in its highly parallelizable nature, as the coin flips for each tuple can occur independently.
For each tuple $j$ in the oblivious output $\mathcal{O}_i$ of the preceding operator $O_i$, the following steps are performed (\Cref{alg:noise-parallel}):

\begin{enumerate}
    \item A weighted coin is tossed.
    \item If the tuple is part of the true output (i.e., $c_i[j] = 1$), its corresponding bit $k_i[j]$ is set to $1$, regardless of the coin toss result.
    \item Else, i.e., the tuple is not part of the true output ( $c_i[j] = 0$), the outcome of the coin toss probabilistically determines its inclusion in the output. A successful toss sets $k_i[j] = 1$, while a failure excludes it by setting $k_i[j] = 0$.
\end{enumerate}

With the success probability $p_i$ sampled from the pre-configured distribution $p_i \sim \mathcal{F}(\theta)$, the output after the \resizer~satisfies $S_i = T_i + E(\eta_i) \leq N_i$, where $E(\eta_i) = p_i \times (N_i-T_i)$, as dictated by the Binomial distribution.

Tuple-level independence enables natural parallelism, greatly reducing the \resizer's performance overhead. This makes the method well-suited for MPC execution (\Cref{subsec:resizer_cost}) by requiring fewer synchronization rounds. Furthermore, its stochastic nature strengthens the privacy guarantees, as discussed in \Cref{subsec:sec_perf}.

\begin{algorithm}[t]
\footnotesize
\caption{Marking of Filler Tuples}
\label{alg:noise-parallel}
\SetKwInput{KwData}{Input}
\SetKwInput{KwResult}{Output}

\KwData{
    \begin{itemize}
        \item Sampled success probability $p_i$.
        \item Oblivious output size $N_i$ and Column $c_i$, indicating true tuples, of the preceding SQL operator $O_i$.
    \end{itemize}
}
\KwResult{Column $k_i$ indicating tuples to retain.}

\BlankLine

\textbf{Step 1: Initialize Data Structures} \\
$k_i \gets []$ \tcp{Initialize an empty column.}
$rand \gets []$ \tcp{Initialize a data structure for sampling random values.}
\BlankLine
\textbf{Step 2: Coin Tossing for Random Selection of Filler Tuples} \\
\For{$j \gets 0$ \KwTo $N_i-1$}{
     $rand[j] \sim U(0, 1)$ \tcp{Flip a coin.}
    $k_i[j] \gets ((rand[j] < p_i) \lor c_i[j])$ \tcp{Compute $k_i[j]$.}
}
\BlankLine
\textbf{Step 3: Return Updated Column} \\
\Return $k_i$
\end{algorithm}

Comparison, logic-or, and sampling uniform random numbers of the tuple marking step (\Cref{alg:noise-parallel}) are directly mapped to their MPC counterparts.
The coin flipping, i.e., sampling from the uniform distribution, can be performed in the pre-processing phase of the MPC protocol, and secret shares of the $N$ random numbers are distributed among the parties.
In the online phase, the parties compute jointly over the secret shares of both the random values and the inputs. 

Since the marking step in \Cref{alg:noise-parallel} exhibits natural parallelism, i.e., its loop rounds are independent of each other, it can benefit from the batching mechanism in MP-SPDZ \footnote{\url{https://mp-spdz.readthedocs.io/en/latest/runtime-options.html}} for improved efficiency. For our setup, we select a batching size of $100,000$.

\subsubsection{Shuffle and Trim}

A linkage attack occurs when an adversary attempts to exploit information from the intermediate results (in our case, $k_i$), thereby inferring sensitive connections between secret shares and the corresponding real database values. For instance, if $k_i$ is revealed without shuffling, the attacker can correlate the positions of true negative matches with false positive matches by leveraging memory from a previous execution. If the attacker can identify all true negative matches, they can also determine the true positive data layout.

To prevent linkage attacks \cite{linkage_attack_eskandarian2021clarion}, shuffling is performed after the \resizer~marked the filler tuples (i.e., column $k$ has been created) and before actually trimming the intermediate results. This ensures that no adversary can link secret shares by observing the outputs of different oblivious operators or repeating the same or similar queries multiple rounds. MP-SPDZ~\cite{keller2020mp-spdz} implements state-of-the-art shuffling \cite{waskman_efficient_secure_shuffle,Asharov22} protocol using the Waksman network~\cite{sasy2023waks_secure_shuffling_sorting}. 

In trim, the computing nodes send their shares of $k$ to reconstruct $k$ and discard tuples with $k[j] = 0$.

\subsubsection{Complexity Analysis}
\label{subsec:complexity}
The computational and communication complexities of the \resizer~steps vary by operation in terms of its input size, $N$, number of oblivious tuples, and $M$, the tuple width in bytes, and output size $S$. 
Tuple marking scales linearly with the number of tuples $N$, yielding $O(N)$ complexity in both dimensions. Although the algorithm theoretically requires $O(N)$ synchronization rounds, the tuple-level independence of parallel tuple marking reduces the cost through the unrolling and batching mechanism used in MP-SPDZ~\cite{keller2020mp-spdz}, thereby improving practical performance under MPC. In the RSS protocol, shuffling~\cite{keller2014efficient_mpspdz_shuffle} is not computationally expensive compared to sorting, with a per-party cost of $O(M N \log N)$ for data of length $M$, and communication cost of $O(N)$. Finally, the trim step combines $O(N)$ and $O(S)$ computation, where $S$ is the size of the trimmed result, but requires only $O(1)$ communication.
In our three-party setting, it is expected that the runtime will be dominated by the communication cost.

\subsubsection{Placement Strategies}

In addition to the flexibility in defining a distribution to sample from for the number of filler tuples to keep, \textsc{\mywrap}~ creates the opportunity to flexibly decide which operators to include a \resizer~after. while clearly beneficial in cases when it leads to significant trimming of intermediate results sizes, depending on the expected selectivity and location of the operator in the query tree, including a \resizer~can even lead to slowdowns. We explore this question in detail in~\Cref{subsec:sec_perf}.

\subsection{User-defined Distributions}
\label{subsec:distr}

The role of this distribution is to dictate the amount of filler tuples $\eta_i$ marked to be kept from trimming. Ideally, $\eta_i$ should not be greater than $N_i - T_i$, i.e., in the range of $[0, N_i - T_i]$. This is inherently enforced through the loop of \Cref{alg:noise-parallel} in \Cref{subsec:noise_add}. By construction, the \resizer~will always maintain true tuples, regardless of the outcome of the coin flip. In \textsc{\mywrap}, the output of the distribution is actually a probability $p_i$ which determines $\eta_i$ as the expected value of flipping a $p_i$ weighted coin $N_i - T_i$ times. 

The shape and parameters of the distribution strongly affect both performance and privacy. The former is public knowledge, so distributions which produce only a single value (e.g., $p_i=0.1$) make inferring the true intermediate result size trivial. A distribution biased toward smaller $p_i$ values results, on average, in fewer filler tuples being retained in $\mathcal{O}_i$, which improves performance. Conversely, a distribution with less bias (i.e., closer to uniform) may result in more filler tuples being retained; however, it increases the uncertainty of intermediate result sizes, thereby making it harder for an attacker to recover the true intermediate size $T_i$, as more observations of $S_i$ values (i.e., more query executions) would be required. In \Cref{sec:security}, we present a metric that computes how many observations of $S_i$ are needed by the attacker to recover $T_i$ under specific conditions and show how different distribution parameters affect the number of observations required to recover $T_i$ in \Cref{subsec:sec_perf}. 

The objective is to select a distribution shape and parameters that jointly satisfy the performance and privacy requirements of data analysts and data owners. Because the distribution must be determined during query planning, its configuration is limited to the information available at that stage, such as the input size and, when available, operator selectivity. This decision can be further guided by our metric in conjunction with an estimate of the system’s performance cost, expressed as the total estimated number of tuples to be processed per query (as shown in \Cref{fig:3join_tc}).  

Overall, the design space of distributions is very large, and it is beyond the scope of this work to automate the process of defining or tuning them for specific workloads. That being said, we provide two example strategies:  general distribution and a distribution that can be used to satisfy $\epsilon$-differential privacy (DP) on intermediate results sizes, that is, the truncated Laplace distribution as deployed in Shrinkwrap~\cite{shrinkwrap_bater2018shrinkwrap}.

\subsubsection{Example Distribution: Beta}
\label{subsec:noisegen_examples}
An intuitive choice for sampling the success probability $p_i$ for the tuple-marking step is the Beta distribution, $Beta(\alpha_i, \beta_i)$. Combined with the Binomial distribution $B(N_i - T_i, p_i)$ representing the tuple marking step, this yields a Beta-Binomial model, $BetaBin(N_i - T_i, \alpha_i, \beta_i)$. This formulation allows $p_i$ to be drawn directly from $[0,1]$ and then applied as the coin-flip probability over $N_i$ tuples. 
For each inserted \resizer, the parameters $\alpha_i$ and $\beta_i$ can be pre-configured independently at the query planning phase. The expected number of filler tuples $E(\eta_i)$ can be computed as $\frac{\alpha_i}{\alpha_i+\beta_i} \times (N_i-T_i)$. 

To improve the overall query performance, the number of (expected) filler tuples $E(\eta_i)$ should be minimized. 
Since $ N_i - T_i$ cannot be controlled at the query planning phase, $\alpha_i$ and $\beta_i$ should be chosen carefully. The larger the difference between $\beta_i$ and $\alpha_i$, with $\beta_i > \alpha_i$, the stronger the tendency toward smaller $p_i$ values and thus smaller $E(\eta_i)$. For instance, a Beta distribution with $\alpha_i = 2$ and $\beta_i = 6$ represents a skewed distribution where the probability $p_i$ is more likely to be closer to 0 than to 1 and the average number of filler tuples $E(\eta_i) = 0.25 \times (N_i - T_i)$. 

\subsubsection{Example Distribution: DP-based} 

To provide $\epsilon$-differential privacy guarantees for the intermediate results sizes, distributions such as the truncated Laplace distribution or the exponential distribution can be deployed.
For example, in Shrinkwrap \cite{shrinkwrap_bater2018shrinkwrap}, truncated Laplace distribution $TLap(\epsilon_i,\delta_i,\Delta c_i)$ is deployed and pre-configured for each operator $O_i$ during the query planning phase: 
\begin{itemize}
    \item Sensitivity ($\Delta c_i$) of operator $O_i$ quantifies how much an SQL operator's output can change when a single database record is modified.
    \item Privacy budget ($\epsilon_i$) reflects the total privacy loss incurred by a mechanism, often expressed using an $(\epsilon_i, \delta_i)$-differential privacy guarantee.
    \item Scale parameter ($b_i$) is derived from the sensitivity and privacy budget as $b_i = \Delta c_i/\epsilon_i$.
    \item Location parameter ($\mu_i$) defines the center of the Laplace distribution and is set to a positive value to favor non-negative samples.
\end{itemize}

Under DP rules, a privacy budget must be maintained, and upon repeated executions of queries, the privacy parameters need to be adjusted accordingly. The effect of this is that, over time, $S_i$ will grow. After exhausting the privacy budget, operators must behave fully oblivious.



DP-based distributions can be deployed with our \textit{tuple marking} approach by computing $p_i$ from $\eta_i$: This is done by trimming $\eta_i$ at runtime to $min(N_i-T_i, \eta_i)$ and dividing by $N_i-T_i$ to ensure the success probability of $p_i \in [0,1]$. The expected number of filler tuples value $E(\eta_i)$ is again $p_i \times (N_i - T_i)$.

\subsection{Proposed Metric: Rounds to Recover}
\label{sec:security}
To better understand the effect of \textsc{\mywrap}~on the privacy of intermediate results \textit{sizes}, we define a conservative metric that computes the \emph{ number of equivalent repetitions of an operator required to recover the operator's true result size $T$} (in short, Rounds to Recover). An equivalent repetition of an operator's execution occurs when, even in different queries, the operator is executed with the same input and yields the same true intermediate result size. 
This definition could be further refined to account for adversarially crafted queries designed to exploit repetitions of an operator, which take slightly different inputs and produce slightly different output sizes. However, such considerations fall outside the scope of this work, as we focus on enterprise settings where data analysts are not assumed to act maliciously.

The metric quantifies how often a pre-configured \resizer~ placed after some operator can execute on a dataset $\mathcal{D}$ before the size of the true intermediate result of the operator has to be considered as revealed. Note that this metric does not quantify the privacy loss of individual tuples in the intermediate results due to deploying \textsc{\mywrap}. 
Assuming a passive attacker (see \Cref{sec:assumption}) collects $r$ observations (samples) of the operator $O$ trimmed output size over $r$ queries: 
$S_1, S_2 \dots S_r$, where $k = 0 \ldots r-1$ and each $S_k = T + \eta_k$ and $\eta_k$ is the amount of filler tuples that is \textit{independently sampled} from a distribution $\mathcal{F}(\mu_\eta, \sigma^2_\eta)$ with a \textit{finite} mean value $\mu_\eta$ and a \textit{finite} variance value $\sigma^2_\eta$, and added to $T$, our secret value. 
We assume the distribution characteristics to be known.

The trimmed output size $S$ follows the same distribution of $\eta$ with a mean value $\mu_s = \mu_\eta + T$ and a variance value $\sigma^2_s = \sigma^2_\eta$. 

After sufficient observations, the attacker can compute the average of observations $\bar{S} = \frac{1}{r} \sum_{k=1}^{r} S_k$ to approximate/estimate the mean value $\mu_s$. That is:
\[
\bar{S} = T + \frac{\sum_{k=1}^{r} \eta_k}{r} = T + \bar{\eta} \approx T + \mu_\eta.
\]
Hence, the attacker can estimate the true output size: 
$T \approx \bar{S} - \mu_\eta.$
Based on the Central Limit Theorem (CLT), $\bar{S}$ follows a normal distribution, with a mean value $\mu_s$ and a variance value $\sigma^2_s/r$, for large enough $r$.
Using confidence intervals \mbox{$
P(\lvert \bar{S} - \mu_s \rvert \leq \text{err}) \geq \alpha$},
we can set an error margin, \mbox{$\text{err} \leq z_{\alpha/2} \times \frac{\sigma_s}{\sqrt{r}}$},
where $(z_{\alpha/2}$ is the z-score corresponding to the desired confidence level. For example, for $\alpha = 99.9\%$, $(z_{\alpha/2} = 3.291)$. 

Therefore, the number of rounds $r$, representing our RtR metric, can be computed as
\begin{equation}
    RtR \geq z^2_{\alpha/2} \times \frac{\sigma^2_s}{\text{err}^2} 
\label{eq:rounds}
\end{equation}


\section{Evaluation}
\label{sec:eval}
We describe our setup and objectives in \Cref{subsec:setup}. Next, we analyze the cost of \resizer~and compare it to SQL operators in \Cref{subsec:resizer_cost}, examine the performance–privacy trade-off in \Cref{subsec:sec_perf}, evaluate the runtime of HealthLnk and TPC-H queries, and finally compare \textsc{\mywrap} to related work in \Cref{subsec:overall_perf}.

\subsection{Setup and Goals}
\label{subsec:setup}
We evaluate \textsc{\mywrap}~on three physical machines, each equipped with 256 GBx2 of RAM and two with Intel(R) Xeon(R) Gold 5220 CPU @ 2.20 GHz, another one with Intel(R) Xeon(R) Gold 5120 CPU @ 2.20 GHz, providing a robust and scalable environment for secure computation. Data for the queries is loaded locally from each machine, and an additional external client starts the execution. We evaluated each query individually over three computing nodes in a local area network (LAN) environment with an average round-trip time (RTT) latency of 0.25 ms, averaging the results across five independent runs.
Overall, our experimental evaluation sets out to provide answers to three sets of questions:

$Q_1$: \emph{How expensive is a \resizer~compared to oblivious SQL operators? How does its runtime scale with more and wider tuples?}~--~We provide answers in \Cref{subsec:resizer_cost} and show that a \resizer~is comparable in runtime cost to the operators and gracefully scales with increasing data sizes.

$Q_2$: \emph{What is the practical implication of the performance--privacy trade-off when revealing intermediate result sizes? How easily can an attacker learn the true intermediate result sizes?} -- This trade-off is explored in~\Cref{subsec:sec_perf}, where we show that there is no silver bullet -- but there is a large space to explore and, with \textsc{\mywrap}, future query optimizers will be able to decide where to use how much protection in order to reach a performance target.

$Q_3$: \emph{How much faster are queries when \textsc{\mywrap}~is used to trim the intermediate result sizes? Are the speedups similar to expectations based on related works?} -- We explore this in \Cref{subsec:overall_perf} using HealthLnk and TPC-H queries. We demonstrate that \textsc{\mywrap}~ enables orders of magnitude faster query execution than fully oblivious and other related work approaches, while maintaining at least the same level of security as other related work that discloses trimmed intermediate result sizes.



\begin{figure}[t]
    \centering
    \begin{subfigure}[b]{0.45\textwidth}
        \small
        \begin{tikzpicture}
        \begin{axis}[
            width=1.0\textwidth,
            height=0.55\textwidth,
            xlabel={Resizer Input Size [Rows]},
            ylabel={Time [seconds]},
            xmode=log,
            ymode=log,
            reverse legend,            
            legend style={at={(0.015,0.85)},anchor=west}
        ]
            
            \addplot[color=orange, mark=square*] coordinates {
                (1e2, 0.01) (1e3, 0.03) (1e4, 0.2) (1e5, 2.3) (1e6, 24.5)
            };
            \addplot[color=blue, mark=o] coordinates {
                (1e2, 0.04) (1e3, 0.56) (1e4, 5.68) (1e5, 60.4) (1e6, 604)
            };
            \addplot[color=green, mark=triangle] coordinates {
                (1e2, 0.17) (1e3, 1.32) (1e4, 7.32) (1e5, 64.5) (1e6, 675)
            };

            \draw[->,shorten >=1mm,shorten <=1mm] (rel axis cs:0.55,0.2) node[right,align=left] {Best performance and \\ highest privacy}  -- (axis cs:7000, 0.12);
            
            \legend{Resizer, Counter-based Resizer, Sort\&Cut}

            \addplot[color=black, mark=none, dash pattern=on 2pt off 2pt on 2pt off 2pt] coordinates {
                (1e2, 0.0025) (1e3, 0.0245) (1e4, 0.245) (1e5, 2.45) (1e6, 24.5)
            } node[pos=0.1,below] {\scriptsize{$O(N)$}};

        \end{axis}
        \end{tikzpicture}
        \caption{Runtime with increasing row count and fixed row width of 16 bytes (plaintext)}
        \label{subfig:a1}
    \end{subfigure}
    \hfill
    \begin{subfigure}[b]{0.45\textwidth}
            \small
        \vspace{1em}
        \begin{tikzpicture}
        \begin{axis}[
            width=1.0\textwidth,
            height=0.55\textwidth,
            xlabel={Number of Columns in Resizer Input (4 bytes per column)},
            ylabel={Time [seconds]},
            xmode=log,
            log basis x=2,
            ymin=0,
            reverse legend,            
            legend style={at={(0.075,0.4)},anchor=west}
        ]
            
            \addplot[color=orange, mark=square*] coordinates {
                (4, 0.18) (8, 0.22) (16, 0.26) (32, 0.36) (64, 0.48)
            };
            \addplot[color=blue, mark=o] coordinates {
                (4, 5.68) (8, 5.88) (16, 6.1) (32, 6.3) (64, 6.5)
            };
            \addplot[color=green, mark=triangle] coordinates {
                (4, 7.12) (8, 7.3) (16, 7.38) (32, 7.56) (64, 7.63)
            };
            \legend{Resizer, Counter-based Resizer, Sort\&Cut, }

            \addplot[color=black, mark=none, dash pattern=on 2pt off 2pt on 2pt off 2pt] coordinates {
                (4, 0.16) (8, 0.24) (16, 0.32) (32, 0.4) (64, 0.48)
            } node[pos=0.5,above] {\scriptsize{$O(log_2(M))$}};

        \end{axis}
        \end{tikzpicture}
        \caption{Runtime with increasing row width and fixed row count of 10000}
        \label{subfig:a2}
    \end{subfigure}
    \caption{\resizer~demonstrates linear scalability with the number of rows and logarithmic scalability with the number of columns. It outperforms the counter-based version and sort\&cut by more than an order of magnitude.
    \label{fig:a1a2}}
\end{figure}

\subsection{Runtime Cost of \resizer}
\label{subsec:resizer_cost}
\textbf{Resizer vs. other approaches.} In \Cref{fig:a1a2}, In \Cref{fig:a1a2}, we examine how the runtime of \resizer~scales with increasing input size, measured by the number and width of tuples.
We compare \resizer~against two alternatives:
(1) A counter-based \resizer~that has the same steps as \resizer~ (mark, shuffle, and trim). For this approach, the number of filler tuples to retain is sampled directly, $\eta_i \sim \mathcal{F}(\theta)$, and $\eta_i$ is assumed to be some value between $[0, N_i]$. 
In counter-based \resizer~'s mark step, the tuples are processed sequentially in order.
For each tuple $j$ of $\mathcal{O}_i$, the following is executed: If the counter has not reached the $\eta_i$ value, or if the tuple belongs to the true output (i.e., $c_i[j] = 1$), its corresponding bit $k_i[j]$ is set to 1. Otherwise, the tuple is excluded, and $k_i[j]$ is set to 0. 
This step is followed by shuffling and trimming.

(2) A sort\&cut approach as presented in \cite{shrinkwrap_bater2018shrinkwrap}. Here, the tuples of $\mathcal{O}_i$ are first sorted in descending order based on column $c_i$, after which some filler tuples are trimmed from the bottom. The number of filler tuples retained in $\mathcal{O}_i$ is determined by sampling the truncated Laplace distribution. 

Note that the selectivity and the number of filler tuples have no effect on the \resizer's runtime since all tuples must be processed by the \resizer.
\Cref{subfig:a1} shows the behavior of \resizer~with increasing number of tuples, where each tuple has a fixed width of $4$ columns, and each column is $4$ bytes in plaintext, resulting in a tuple width of $4 \times 4 = 16$ bytes. We compare the runtime of \resizer~with the counter-based \resizer~ and our implementation of the sort\&cut method~\cite{shrinkwrap_bater2018shrinkwrap} in MP-SPDZ.

As expected, \resizer~ runtime scales linearly with the number of rows in its input, still an order of magnitude faster than the counter-based \resizer. Although the counter-based \resizer~ is simple to implement, its inherently sequential nature renders it inefficient for MPC execution. It requires a secure counter to be checked in each iteration, to determine whether the count of filler tuples marked so far has reached the limit $\eta_i$, thereby creating a loop dependency. In contrast, for the \resizer~ each iteration is independent, enabling parallelism and thus more efficient execution.

Furthermore, as illustrated in \Cref{subfig:a1}, both \resizer~and counter-based \resizer~outperform the sorting-based solution. This improvement arises because secure shuffling is less computationally expensive than secure sorting \cite{shrinkwrap_bater2018shrinkwrap}. 

We also conducted an experiment to show that the width of tuples plays a less critical role in the runtime of the \resizer. We fixed the number of tuples at $ N=10,000$. As the table width increases from 2 to 64 columns (with $10,000$ rows fixed), the \resizer~runtime rises gradually, exhibiting a logarithmic growth. Tuple width does not affect the mark or trim steps of the \resizer, both steps operate purely row-wise, acting on the $c_i$ and $k_i$ columns. However, the shuffle step involves a copy operation, which can be slightly impacted by the width of the tuples being copied. As shown in \Cref{subfig:a2}, increasing the number of columns results in only a sublinear increase in cost.
Overall, the runtime of \resizer~is both the lowest compared to the other variants.

\textbf{Combining Oblivious Operators with Resizer.} 
In this experiment, we compare the runtime overhead of \resizer~to other database operators under different intermediate result sizes (before trimming). In the context of a single operator followed by a \resizer, the number of rows that are being trimmed away plays no role in the runtime of either the operator or the \resizer. Its effect is only visible for the next operator, with less input data to process. For this reason, in~\Cref{fig:a3_1} and~\Cref{fig:a3_2}, we do not show operator selectivity or the count of filler tuples.

\begin{figure}[t]
    \centering
        \small
        \begin{tikzpicture}
        \begin{axis}[
            width=0.65\columnwidth,
            height=0.55\columnwidth,
            xlabel={Oblivious Operator Result Size [Rows]},
            ylabel={Time [seconds]},
            xmode=log,
            ymode=log,
            legend style={at={(1.05,0.5)},anchor=west},
            grid=major
    ]

            \addplot[
                mark=triangle*, 
                color=green
            ] coordinates {(100, 0.32) (1000, 1.54) (10000, 8.9) (100000, 79.2) (1000000, 727)};
            \addlegendentry{GroupBy+Resizer}
            
                \addplot[
                mark=triangle, 
                color=green
            ] coordinates {(100, 0.29) (1000, 1.4) (10000, 8.32) (100000, 64.4) (1000000, 678.1)};
            \addlegendentry{GroupBy}

         \addplot[
                mark=*, 
                color=blue               
            ] coordinates {(100, 0.04) (1000, 0.12) (10000, 0.64) (100000, 4.5) (1000000, 38.9)};
            \addlegendentry{Join$_B$+Resizer}
            
            \addplot[
                mark=o, 
                color=blue                
            ] coordinates {(100, 0.02) (1000, 0.07) (10000, 0.22) (100000, 1.55) (1000000, 8.9)};
            \addlegendentry{Join$_B$}

                  \addplot[
                mark=square*, 
                color=red 
            ] coordinates {(100, 0.01) (1000, 0.04) (10000, 0.26) (100000, 3.24) (1000000, 33.8)};
            \addlegendentry{Filter$_1$+Resizer}

            \addplot[
                mark=square, 
                color=red
            ] coordinates {(100, 0.005) (1000, 0.01) (10000, 0.08) (100000, 1.03) (1000000, 8.14)};
            \addlegendentry{Filter$_1$}

    \end{axis}
    \end{tikzpicture}
    \caption{Adding \resizer~after an operator increases runtime, but reducing the output size will speed up the next operator.}
    \label{fig:a3_1}
\end{figure}







\Cref{fig:a3_1} shows the runtime of oblivious operators with and without a \resizer. We focused here on (1) a filter with one equality condition (Filter$_1$), (2) a join that has two balanced tables as input, each has a size equal to the square root of the output size (Join$_B$), and (3) a group by operator that groups based on one column. Thanks to the linear runtime of \resizer, which is independent of the type of operator executing before it, the results in \Cref{fig:a3_1} are predictable, as performance does not degrade significantly.

In \Cref{fig:a3_2}, we show how expensive a \resizer~ is relative to the actual operators, with a fixed 1M intermediate result size (i.e., the oblivious output of the operator and the input to the \resizer~has 1M rows). We compared Filter with one equality condition (Filter$_1$) and four equality conditions (Filter$_4$), a balanced join (Join$_B$) with two input tables of size $sqrt(N)$, an unbalanced join (Join$_S$) of $1:N$, and a group by. As seen in \Cref{fig:a3_2}, we tested each step of \resizer~ separately. The mark step of \resizer~ is more expensive than Filter$_1$ but cheaper than the GroupBy. 
This is because for each tuple, the \resizer~ needs to conduct an online comparison and a logical OR gate over secret shares. However, Filter$_1$ only requires one equality check. Similarly, if the operator is more complex, such as GroupBy (which includes sorting as a pre-operation), the mark step of \resizer~will seem relatively cheap. The shuffle step is performed in constant rounds, with the final trim step being cheaper than the previous operations.

\begin{figure}[t]
    \centering
        \small
      \begin{tikzpicture}          
        \pgfplotstableread{
        type	       oblivop    resizer      relobliv     relresizer  relnoiseadding  relsecureshuffling relrevealandcopy  
        $Filter_{1}$	        8.14        25.66       23.84	    75.91715976 66.49 7.62 2.1
        $Filter_{4}$         20.42           24.88           44.18        52.8 49.11 5.19 1.51
        $Join_{B}$   8.9        30           24.51     77.12082262 65.56 7.44 2.48
        $Join_{S}$   8.7        30           24.5    77.51937984  65.6 6.76 3.1 
        $GroupBy$         691.1       27          96.1     3.7  3.4 0.34 0.097
        }\datatable
        
        \begin{axis}[
          legend style={at={(1.4,1)},anchor=north east},
          width=0.35\textwidth,
          height=0.27\textwidth,      
          grid=major,
           ybar stacked,
          bar width=15pt,  
          legend reversed,  
          ylabel={Relative runtime for 1M rows [\%]},
            xticklabels from table={\datatable}{type},
            xtick=data,
            x tick label style={rotate=30,anchor=east},  
            ]
        
        \addplot [blue!90, pattern=north west lines, pattern color=blue!50] table [x expr=\coordindex, y=relobliv] {\datatable};
        
        \addplot [red!90, fill=red!20] table [x expr=\coordindex, y=relnoiseadding] {\datatable};
        \addplot [green!90, fill=green!20] table [x expr=\coordindex, y=relsecureshuffling] {\datatable};
        \addplot [black!90, fill=black!20] table [x expr=\coordindex, y=relrevealandcopy] {\datatable};
        
        \legend{Operator, Mark, Shuffle, Trim}
        
        \end{axis}
        \end{tikzpicture}   
    \caption{Depending on the oblivious operator, \resizer~ could be relatively cheap compared to the operator itself. Filter$_1$ is a filter with 1 equality condition,  Filter$_4$ is a filter with 4 equality conditions, Join$_B$ has two balanced tables, each with a size equal to the square root of the output size, as input, and Join$_{S}$ has two unbalanced tables as input.}
    \label{fig:a3_2}
\end{figure}

\subsection{Privacy vs. Performance}
\label{subsec:sec_perf}
In this evaluation, we use the Rounds to Recover (RtR) in \Cref{eq:rounds} derived in \Cref{sec:security}. RtR shows how often an attacker needs to observe the MPC execution of equivalent repetitions to recover the true intermediate size $T_i$ of an operator $O_i$, within a given error margin when \textsc{\mywrap}~is deployed. Recall $RtR \geq z^2_{\alpha/2} \times \sigma^2_s / err^2$. In our experiments, we fix $z_{\alpha/2} = 3.291$, i.e., to compute RtR that achieves a confidence level of $99.9\%$, and examine the effect of different distributions, represented by their variances and error margins. Furthermore, we examine the performance--privacy trade-off.

\begin{figure}[t]
    \centering
    \begin{subfigure}[b]{0.49\columnwidth}
        \footnotesize
        \begin{tikzpicture}
        \begin{axis}[
                width=\linewidth, height=4cm,
                xlabel={N},
                ylabel={Rounds to recover T$\pm$1 (RtR)},
                grid=major,
                 ymode=log,
                xmode=log,
                ymin=10,
                ymax=1000,
                legend style={at={(0.45,1.55)}, anchor=north, legend columns=1},
            ]
            
            \addplot[blue, mark=o] coordinates {
            (100, 87) (20506, 87) (40912, 87) (61318, 87) (81724, 87) (102130, 87) (122536, 87) (142942, 87) (163348, 87) (183755, 87) (204161, 87) (224567, 87) (244973, 87) (265379, 87) (285785, 87) (306191, 87) (326597, 87) (347004, 87) (367410, 87) (387816, 87) (408222, 87) (428628, 87) (449034, 87) (469440, 87) (489846, 87) (510253, 87) (530659, 87) (551065, 87) (571471, 87) (591877, 87) (612283, 87) (632689, 87) (653095, 87) (673502, 87) (693908, 87) (714314, 87) (734720, 87) (755126, 87) (775532, 87) (795938, 87) (816344, 87) (836751, 87) (857157, 87) (877563, 87) (897969, 87) (918375, 87) (938781, 87) (959187, 87) (979593, 87) (1000000, 87)   
            };
            \addlegendentry{Counter-based Resizer}

            \addplot[red, mark=triangle] coordinates {
            (100, 248) (20506, 291) (40912, 291) (61318, 291) (81724, 291) (102130, 291) (122536, 291) (142942, 291) (163348, 291) (183755, 291) (204161, 291) (224567, 291) (244973, 291) (265379, 291) (285785, 291) (306191, 291) (326597, 291) (347004, 291) (367410, 291) (387816, 291) (408222, 291) (428628, 291) (449034, 291) (469440, 291) (489846, 291) (510253, 291) (530659, 291) (551065, 291) (571471, 291) (591877, 291) (612283, 291) (632689, 291) (653095, 291) (673502, 291) (693908, 291) (714314, 291) (734720, 291) (755126, 291) (775532, 291) (795938, 291) (816344, 291) (836751, 291) (857157, 291) (877563, 291) (897969, 291) (918375, 291) (938781, 291) (959187, 291) (979593, 291) (1000000, 291)
            };
            \addlegendentry{Resizer, $T = 0.1N$}
                
            \addplot[green, mark=square] coordinates {  (100, 213) (20506, 291) (40912, 291) (61318, 291) (81724, 291) (102130, 291) (122536, 291) (142942, 291) (163348, 291) (183755, 291) (204161, 291) (224567, 291) (244973, 291) (265379, 291) (285785, 291) (306191, 291) (326597, 291) (347004, 291) (367410, 291) (387816, 291) (408222, 291) (428628, 291) (449034, 291) (469440, 291) (489846, 291) (510253, 291) (530659, 291) (551065, 291) (571471, 291) (591877, 291) (612283, 291) (632689, 291) (653095, 291) (673502, 291) (693908, 291) (714314, 291) (734720, 291) (755126, 291) (775532, 291) (795938, 291) (816344, 291) (836751, 291) (857157, 291) (877563, 291) (897969, 291) (918375, 291) (938781, 291) (959187, 291) (979593, 291) (1000000, 291)               
            };
            \addlegendentry{Resizer, $T = 0.5N$}
            
        \end{axis}
        \end{tikzpicture}
        \caption{RtR for TLap($\Delta c = 1$)}
        \label{subfig:c1_1}
    \end{subfigure}
    \hfill
    \begin{subfigure}[b]{0.49\columnwidth}
            \footnotesize
        \vspace{1em}
        \begin{tikzpicture}
            \begin{axis}[
                width=\linewidth, height=4cm,
                xlabel={N},
                ylabel={Rounds to recover T$\pm$1 (RtR)},
                grid=major,
                ymode=log,
                xmode=log,
                ymin=1000,
                ymax=1000000000,
                legend style={at={(0.45,1.55)}, anchor=north, legend columns=1},
            ]
            
        \addplot[blue, mark=o] coordinates {
        (100, 8665) (20506, 1776752) (40912, 3544839) (61318, 5312926) (81724, 7081013) (102130, 8849100) (122536, 10617187) (142942, 12385274) (163348, 14153361) (183755, 15921535) (204161, 17689622) (224567, 19457709) (244973, 21225796) (265379, 22993883) (285785, 24761970) (306191, 26530057) (326597, 28298144) (347004, 30066318) (367410, 31834405) (387816, 33602492) (408222, 35370579) (428628, 37138666) (449034, 38906753) (469440, 40674840) (489846, 42442927) (510253, 44211100) (530659, 45979187) (551065, 47747274) (571471, 49515361) (591877, 51283448) (612283, 53051535) (632689, 54819622) (653095, 56587709) (673502, 58355883) (693908, 60123970) (714314, 61892057) (734720, 63660144) (755126, 65428231) (775532, 67196318) (795938, 68964405) (816344, 70732492) (836751, 72500666) (857157, 74268753) (877563, 76036840) (897969, 77804927) (918375, 79573014) (938781, 81341101) (959187, 83109188) (979593, 84877275) (1000000, 86645448) 
        };
        \addlegendentry{Counter-based Resizer}

        \addplot[red, mark=triangle] coordinates {
        (100, 6152) (20506, 1802234) (40912, 3582744) (61318, 5360372) (81724, 7136504) (102130, 8911680) (122536, 10686177) (142942, 12460159) (163348, 14233733) (183755, 16007060) (204161, 17780022) (224567, 19552745) (244973, 21325263) (265379, 23097599) (285785, 24869775) (306191, 26641807) (326597, 28413710) (347004, 30185582) (367410, 31957260) (387816, 33728840) (408222, 35500329) (428628, 37271734) (449034, 39043061) (469440, 40814315) (489846, 42585501) (510253, 44356710) (530659, 46127772) (551065, 47898777) (571471, 49669729) (591877, 51440630) (612283, 53211482) (632689, 54982290) (653095, 56753053) (673502, 58523862) (693908, 60294545) (714314, 62065189) (734720, 63835798) (755126, 65606371) (775532, 67376912) (795938, 69147420) (816344, 70917897) (836751, 72688431) (857157, 74458850) (877563, 76229242) (897969, 77999607) (918375, 79769946) (938781, 81540260) (959187, 83310550) (979593, 85080817) (1000000, 86851147)
        };
        \addlegendentry{Resizer, $T = 0.1N$}
                
        \addplot[green, mark=square] coordinates { (100, 2462) (20506, 1798527) (40912, 3579037) (61318, 5356664) (81724, 7132797) (102130, 8907973) (122536, 10682469) (142942, 12456451) (163348, 14230025) (183755, 16003353) (204161, 17776314) (224567, 19549038) (244973, 21321555) (265379, 23093891) (285785, 24866067) (306191, 26638099) (326597, 28410002) (347004, 30181874) (367410, 31953552) (387816, 33725132) (408222, 35496621) (428628, 37268026) (449034, 39039353) (469440, 40810607) (489846, 42581793) (510253, 44353002) (530659, 46124064)(551065, 47895069) (571471, 49666021) (591877, 51436922) (612283, 53207775) (632689, 54978582) (653095, 56749345) (673502, 58520154) (693908, 60290837) (714314, 62061482) (734720, 63832090) (755126, 65602664) (775532, 67373204) (795938, 69143712) (816344, 70914189) (836751, 72684724) (857157, 74455143) (877563, 76225534) (897969, 77995899) (918375, 79766238) (938781, 81536552) (959187, 83306842) (979593, 85077109) (1000000, 86847440)  };
        \addlegendentry{Resizer, $T = 0.5N$}

        \end{axis}
        \end{tikzpicture}
        \caption{RtR for TLap($\Delta c = \sqrt N$)}
        \label{subfig:c1_2}
    \end{subfigure}
    \caption{Coin tossing-based Resizer (in red and green) mostly outperforms counter-based Resizer with $T=0.1N$ (in blue). More observations of the operator's noisy output size $S$ are needed to recover the true output size $T$ with an error margin of 1 tuple. Both figures deploy a truncated Laplace distribution from \cite{shrinkwrap_bater2018shrinkwrap} $TLap(\epsilon = 0.5, \delta = 0.00005)$. The left-side figure has a narrower distribution with $\Delta c = 1$ and a scale $b =\frac{1}{\epsilon} = 2$, whereas the right-side figure has a wider distribution with $\Delta c = \sqrt N$ a scale $b = \frac{\sqrt N}{\epsilon} = 2\sqrt N$.}
    \label{fig:c1}
\end{figure}

\textbf{\resizer~ vs. Counter-based Resizer.} In this experiment, we evaluate the impact of coin tossing-based mark step on RtR. We compare the coin-tossing-based mark with the counter-based mark step \Cref{subsec:resizer_cost}. We use the $\epsilon$-differential truncated Laplace distribution $TLap(\epsilon = 0.5, \delta = 0.00005)$ on the range $[0, \infty)$~\cite{shrinkwrap_bater2018shrinkwrap} and fix the error margin to 1 
 -- in other words, the attacker ``wins'' if $T \pm 1$ can be determined with $99.9\%$ confidence level. 

When using the Counter-based resizer, $\eta$ will be sampled directly from $TLap(\epsilon = 0.5, \delta = 0.00005)$. In case of coin-tossing, $p$ is computed as $\ min(eta, N) / N$ (with $\eta$ sampled as mentioned above). Using the law of variance, we compute the variance of the combined truncated Laplace and Binomial distributions for two values of $T$ ($T = 0.1 \times N$ and $T = 0.5 \times N$). In addition, we test under different sensitivities ($\Delta c = 1$ and $\Delta c = \sqrt N$), which affect the shape of the distribution (narrow or wide).

For the TLap distribution with a low scale $b = 2$ ($\Delta c = 1$), \Cref{subfig:c1_1} demonstrates that the coin tossing-based mark results in more rounds RtR than the counter-based mark, even for larger $T$ values. Counter-based mark can only achieve comparable privacy guarantees when combined with a TLap distribution of a higher scale $b = 2\sqrt{N}$ ($\Delta c = \sqrt N$) as shown in \Cref{subfig:c1_2}.

\textbf{Impact of Distribution.}
In this experiment, we use \resizer~ with two different distributions: the truncated Laplace distribution, as defined above, and the Beta distribution $B(\alpha=2, \beta=6$, described in \Cref{subsec:distr} and utilize an error margin of $T \pm 1$, our results in \Cref{subfig:c2_1} indicate that Beta-binomial distribution achieves more RtR compared to truncated Laplace, even with high scale value $b = 2\sqrt N$.

Larger values of $r$ affect performance. To analyze the trade-offs, we evaluate the runtime of \resizer~with the different distributions: $TLap(\epsilon=0.5, \delta=0.00005, \Delta c=1000)$ and $Beta(\alpha=2, \beta=6)$  under the same workload as in \Cref{fig:resizer-placement} (right-side), with $N=1M$ and $T=10\% N$, for the operation sequence Join$_{B}$ $\rightarrow$ \resizer $\rightarrow$ OrderBy. \resizer~ configured with $TLap(0.5, 0.00005, 1000)$, resulting in an average count of filler tuples of approx. $2\% N$, achieved a runtime of $104$s. In comparison, using $Beta(2,6)$, which introduces an average count of filler tuples of $25\% (N-T) = 22.5\%N$, resulted in a runtime of $236$s. 

\begin{figure}[t]
    \centering
    \begin{subfigure}[b]{0.48\columnwidth}
        \footnotesize
        \begin{tikzpicture}
            \begin{axis}[
                width=\linewidth, height=4cm,
                xlabel={N},
                ylabel={Rounds to recover $T\pm1$ (RtR)},
                grid=major,
                ymode=log,
                xmode=log,
                ymin=0.5,
                legend style={at={(-0.2,1.35)}, anchor=west, legend columns=1},
                reverse legend
            ]
            \addplot[blue, mark=o] coordinates {
                (100, 250) (20506, 291) (40912, 291) (61318, 291) (81724, 291) (102130, 291) (122536, 291) (142942, 291) (163348, 291) (183755, 291) (204161, 291) (224567, 291) (244973, 291) (265379, 291) (285785, 291) (306191, 291) (326597, 291) (347004, 291) (367410, 291) (387816, 291) (408222, 291) (428628, 291) (449034, 291) (469440, 291) (489846, 291) (510253, 291) (530659, 291) (551065, 291) (571471, 291) (591877, 291) (612283, 291) (632689, 291) (653095, 291) (673502, 291) (693908, 291) (714314, 291) (734720, 291) (755126, 291) (775532, 291) (795938, 291) (816344, 291) (836751, 291) (857157, 291) (877563, 291) (897969, 291) (918375, 291) (938781, 291) (959187, 291) (979593, 291) (1000000, 291)
            };
            \addlegendentry{TLap($\Delta c = 1$)}

            \addplot[red, mark=triangle] coordinates {
                (100, 6394) (20506, 1802477) (40912, 3582988) (61318, 5360616) (81724, 7136748) (102130, 8911924) (122536, 10686421) (142942, 12460403) (163348, 14233977) (183755, 16007304) (204161, 17780266) (224567, 19552989) (244973, 21325507) (265379, 23097843) (285785, 24870019) (306191, 26642051) (326597, 28413954) (347004, 30185826) (367410, 31957504) (387816, 33729084) (408222, 35500573) (428628, 37271978) (449034, 39043305) (469440, 40814559) (489846, 42585745) (510253, 44356954) (530659, 46128015) (551065, 47899021) (571471, 49669972) (591877, 51440873) (612283, 53211726) (632689, 54982534) (653095, 56753297) (673502, 58524106) (693908, 60294789) (714314, 62065433) (734720, 63836042) (755126, 65606615) (775532, 67377156) (795938, 69147664) (816344, 70918141) (836751, 72688675) (857157, 74459094) (877563, 76229486) (897969, 77999850) (918375, 79770190) (938781, 81540504) (959187, 83310794) (979593, 85081060) (1000000, 86851391)
            };
            \addlegendentry{TLap($\Delta c = \sqrt N$)}
                
            \addplot[green, mark=square] coordinates { (100, 2208) (20506, 85664712) (40912, 340920050) (61318, 765768221) (81724, 1360209227) (102130, 2124243068) (122536, 3057869742) (142942, 4161089251) (163348, 5433901594) (183755, 6876381612) (204161, 8488387934) (224567, 10269987091) (244973, 12221179081) (265379, 14341963906) (285785, 16632341565) (306191, 19092312058) (326597, 21721875386) (347004, 24521172876) (367410, 27489930183) (387816, 30628280324) (408222, 33936223299) (428628, 37413759109) (449034, 41060887752) (469440, 44877609230) (489846, 48863923542) (510253, 53020038505) (530659, 57345546796) (551065, 61840647922) (571471, 66505341882) (591877, 71339628676) (612283, 76343508304) (632689, 81516980766) (653095, 86860046063) (673502, 92372978498) (693908, 98055237774) (714314, 103907089884) (734720, 109928534828) (755126, 116119572607) (775532, 122480203220) (795938, 129010426667) (816344, 135710242948) (836751, 142579992855) (857157, 149619003116) (877563, 156827606211) (897969, 164205802139) (918375, 171753590903) (938781, 179470972500) (959187, 187357946931) (979593, 195414514197) (1000000, 203641081577)         
            };
            \addlegendentry{Beta($\alpha=2,\beta=6$)}
        \end{axis}
        \end{tikzpicture}
        \caption{RtR with $err =1$}
        \label{subfig:c2_1}
    \end{subfigure}
    \hfill
    \begin{subfigure}[b]{0.48\columnwidth}
            \footnotesize
        \vspace{1em}
        \begin{tikzpicture}
            \begin{axis}[
                width=\linewidth, height=4cm,
                xlabel={N},
                ylabel={Rounds to recover $T\pm1\%N$ (RtR)},
                grid=major,
                 ymode=log,
                xmode=log,
                ymin=0.5,
                 legend style={at={(-0.2,1.35)}, anchor=west, legend columns=1},     
                 reverse legend
                ]
                \addplot[blue, mark=o] coordinates {
                 (100, 250) (20506, 1) (40912, 1) (61318, 1) (81724, 1) (102130, 1) (122536, 1) (142942, 1) (163348, 1) (183755, 1) (204161, 1) (224567, 1) (244973, 1) (265379, 1) (285785, 1) (306191, 1) (326597, 1) (347004, 1) (367410, 1) (387816, 1) (408222, 1) (428628, 1) (449034, 1) (469440, 1) (489846, 1) (510253, 1) (530659, 1) (551065, 1) (571471, 1) (591877, 1) (612283, 1) (632689, 1) (653095, 1) (673502, 1) (693908, 1) (714314, 1) (734720, 1) (755126, 1) (775532, 1) (795938, 1) (816344, 1) (836751, 1) (857157, 1) (877563, 1) (897969, 1) (918375, 1) (938781, 1) (959187, 1) (979593, 1) (1000000, 1)
                };
                \addlegendentry{TLap($\Delta c = 1$)}

                \addplot[red, mark=triangle] coordinates {
                 (100, 6394) (20506, 43) (40912, 22) (61318, 15) (81724, 11) (102130, 9) (122536, 8) (142942, 7) (163348, 6) (183755, 5) (204161, 5) (224567, 4) (244973, 4) (265379, 4) (285785, 4) (306191, 3) (326597, 3) (347004, 3) (367410, 3) (387816, 3) (408222, 3) (428628, 3) (449034, 2) (469440, 2) (489846, 2) (510253, 2) (530659, 2) (551065, 2) (571471, 2) (591877, 2) (612283, 2) (632689, 2) (653095, 2) (673502, 2) (693908, 2) (714314, 2) (734720, 2) (755126, 2) (775532, 2) (795938, 2) (816344, 2) (836751, 2) (857157, 2) (877563, 1) (897969, 1) (918375, 1) (938781, 1) (959187, 1) (979593, 1) (1000000, 1)
                };
                \addlegendentry{TLap($\Delta c = \sqrt N$)}
                
                \addplot[green, mark=square] coordinates {  (100, 2208) (20506, 2038) (40912, 2037) (61318, 2037) (81724, 2037) (102130, 2037) (122536, 2037) (142942, 2037) (163348, 2037) (183755, 2037) (204161, 2037) (224567, 2037) (244973, 2037) (265379, 2037) (285785, 2037) (306191, 2037) (326597, 2037) (347004, 2037) (367410, 2037) (387816, 2037) (408222, 2037) (428628, 2037) (449034, 2037) (469440, 2037) (489846, 2037) (510253, 2037) (530659, 2037) (551065, 2037) (571471, 2037) (591877, 2037) (612283, 2037) (632689, 2037) (653095, 2037) (673502, 2037) (693908, 2037) (714314, 2037) (734720, 2037) (755126, 2037) (775532, 2037) (795938, 2037) (816344, 2037) (836751, 2037) (857157, 2037) (877563, 2037) (897969, 2037) (918375, 2037) (938781, 2037) (959187, 2037) (979593, 2037) (1000000, 2037)               
                };
                \addlegendentry{Beta($\alpha=2,\beta=6$)}  
            \end{axis}
        \end{tikzpicture}
        \caption{RtR with $err =1\%N$}
        \label{subfig:c2_2}
    \end{subfigure}
    \caption{RtR highly depends on the error margin accepted by the attacker. Combining \resizer~ with Beta distribution (green plots) results in more observations required to recover $T$ even with the attacker tolerating accuracy (by allowing a higher error margin in \Cref{eq:rounds}). Note that all plots use $T = 0.05 N$. For a narrow truncated Laplace distribution ($\Delta c = 1$), RtR = 1, i.e., one observation, for small $T$ values.}
    \label{fig:c2}
\end{figure}

We analyze the impact of the error margin on the number of rounds an attacker needs for the two distributions when combined \resizer. Relaxing the error margin allows for an informal examination of the potential extent of information leakage on $T$. Specifically, we investigate how many rounds an attacker would need to guess $T$ within a specified range relative to $N$. Our findings in \Cref{subfig:c2_2} demonstrate that relaxing the accuracy requirement for $T$ can dramatically reduce the number of rounds. For instance, with small $T$ values (e.g., $T = 5\% \times N$), using a narrow distribution ($b = 2$) and permitting a recovery accuracy of $T \pm 1\% \times N$ reduces the number of rounds to $RtR = 1$ (i.e., the attacker needs to observe only a single repetition to discover $T = 5\% \times N$). In contrast, wider distributions can still offer acceptable guarantees, but the expected runtime will be significantly larger, as fewer tuples are trimmed on average. 
We believe the RtR metric is a crucial step towards enabling the query optimizer to consider the privacy implications of different distributions and balance them with their performance implications.



\textbf{Resizer Placement -- Micro-benchmarks.}
Based on the results in \Cref{fig:a3_2}, it is clear that a \resizer~incurs an additional runtime cost, which must be balanced against the benefits of data reduction as the query tree is traversed. In the previous experiment, we placed a \resizer~after each internal operator, but in the future, a query optimizer should decide on a per-operator basis, based on its expected selectivity and the magnitude of expected filler tuples, whether a \resizer~is worth inserting or not. To show what \textit{cost functions} an optimizer might use, we depict the performance of two recurring operator combinations: a join followed by a filter (Join$_B$ $\rightarrow$ Filter$_1$), and a join followed by an order by (Join$_B$ $\rightarrow$ OrderBy). We test the overall runtime with and without a \resizer~ after the join in both combinations.

\begin{figure}[t]
    \small
    \centering
    \begin{minipage}[b]{0.45\linewidth}
        \centering
        \begin{tikzpicture}
            \begin{axis}[
                width=\linewidth, height=3.3cm,
                xlabel={Join Selectivity},
                ylabel={Execution Time [s]},
                grid=major,
                ymin=0,
                xmax=1,
                legend style={font=\footnotesize, at={(0.35,1.05)}, anchor=south, legend columns=1},
                ]
                \addplot[blue, mark=o] coordinates {
                    (0.01, 41.4) (0.1, 42) (0.2, 42.84) (0.3, 43.6) (0.4, 44.5) (0.5, 46)
                    (0.6, 47.2) (0.7, 48.1) (0.8, 49) (0.9, 49.8)
                };
                \addlegendentry{Join$_B$+\resizer+Filter$_1$}

                \addplot[red] coordinates {
                    (0.01, 33) (0.1, 33) (0.2, 33) (0.3, 33) (0.4, 33) (0.5, 33)
                    (0.6, 33) (0.7, 33) (0.8, 33) (0.9, 33)
                };
                \addlegendentry{Join$_B$+Filter$_1$}
            \end{axis}
        \end{tikzpicture}
        \label{fig:subplot1}
    \end{minipage}%
    \hspace{0.05\linewidth}
    \begin{minipage}[b]{0.45\linewidth}
        \centering
        \begin{tikzpicture}
            \begin{axis}[
                width=\linewidth, height=3.3cm,
                xlabel={Join Selectivity},
                ylabel={Execution Time [s]},
                grid=major,
                ymin=0,
                xmax=1,
                legend style={font=\footnotesize, at={(0.36,1.05)}, anchor=south, legend columns=1},
                ]
                \addplot[blue, mark=o] coordinates {
                    (0.01, 113.7) (0.1, 149.72) (0.2, 207.9) (0.3, 258.2) (0.4, 319.1) (0.5, 396.9)
                    (0.6, 449.4) (0.7, 504.3) (0.8, 570) (0.9, 597.5)
                };
                \addlegendentry{Join$_B$+\resizer+OrderBy}

                \addplot[red] coordinates {
                    (0.01, 566) (0.1, 566) (0.2, 566) (0.3, 566) (0.4, 566) (0.5, 566)
                    (0.6, 566) (0.7, 566) (0.8, 566) (0.9, 566)
                };
                \addlegendentry{Join$_B$+OrderBy}
            \end{axis}
        \end{tikzpicture}
        \label{fig:subplot2}
    \end{minipage}
    \vspace{-3ex}
    \caption{The left figure shows that inserting \resizer~(that adds 10\% of the FO-Join output size) between Join and Filter is not beneficial, whereas the right figure shows that between Join and OrderBy it will be almost always beneficial.}
    \vspace{-3ex}
    \label{fig:resizer-placement}
\end{figure}

In \Cref{fig:resizer-placement}, we show the performance of the query snippets, assuming different selectivity of the join. Interestingly, in \Cref{fig:resizer-placement} (left-side), the runtime with \resizer~ is always higher than without it. Hence, placing the \resizer~ before the filter, with the filter as the last operator in the query, would actually slow it down. For the case in \Cref{fig:resizer-placement} (right-side), the opposite is true, and inserting the \resizer~ speeds up execution except for selectivity $>85\%$.

In fully oblivious execution, 
the query plans have little space for optimization. With \textsc{\mywrap}, query optimization is possible. As this example shows, it is straightforward to construct cost functions for future query optimizers to inform their decisions.

\textbf{Effect of \resizer~Placement in a TPC-H Query.}
In this experiment, we evaluate the impact of different placement rules for \resizer~on both the runtime and the Rounds to Recovery (RtR). The three rules are: insert a \resizer~after all intermediate operators, only after joins, or only after GroupBys (plotted as lines with marks on \Cref{fig:privacy_vs_runtime_points}). 
We assume the \resizer s are configured to receive $p$ values using the Beta distribution. 
For datasets with $15\%$ selectivity, we used four configurations: $5\%$, $10\%$, $20\%$, and $30\%$ of input size, as the average count of filler tuples. For $85\%$ selectivity, we used: $5\%$, $10\%$, and $15\%$ of the input size as the average count of filler tuples. Although the relative RtR behavior is valid across distributions, it is essential to note that the choice of user-defined distribution has a significant impact on the absolute value and, consequently, the real-world security of the system.

We evaluate the runtime and the RtR metric of these placements on TPC-H Q3, with data modified such that all operators have either around 15\% or 85\% selectivity -- to show the selectivity spectrum on the same query tree. The reported RtR value reflects the minimum among RtR values of all operators in Q3 (e.g., the weakest link in the query plan).

The red star on \Cref{fig:privacy_vs_runtime_points} denotes the execution of Q3 without any \resizer~insertions and with all intermediate result sizes are fully revealed. This configuration yields the best performance but no protection, as a single observation is sufficient to recover the true sizes of the intermediate results of all operators (i.e., $RtR = 1$). In contrast, the red bar represents the execution time of Q3 in fully oblivious mode, where no \resizer~is used. This setup offers the highest level of privacy (i.e., $RtR \to \infty$), and is depicted as a bar to indicate that any strategy with a higher execution time (i.e., that crosses this bar) is considered impractical.

\begin{figure}
\centering
\begin{tikzpicture}
\scriptsize
\begin{axis}[
    width=5.5cm, height=4.3cm,
    xlabel={Minimum RtR across operators},
    ylabel={Query Runtime [s]},
    xmin=1, xmax=100000000000,
    ymin=0, ymax=2000,
    axis lines=middle,
    xmode = log,
    axis line style={->},
    xlabel near ticks,
    ylabel near ticks,
    grid=both,
    major grid style={dashed,gray!30},
    legend style={at={(1.9,0.01)}, anchor=south east},
    every axis legend/.append style={cells={anchor=west}}
]

\addplot[
    green,
    thick,
    mark=square,
    mark size=2pt
]
coordinates {
    (158, 3.07)
    (185, 9.4)
    (192, 24.3)
    (196, 67.3)
};
\addlegendentry{After all (s=15\%)}

\addplot[
    green,
    thick,
    mark=square*,
    mark size=2pt
]
coordinates {
    (83,928.2)
    (158, 1264.95)
    (180, 1763)
};
\addlegendentry{After all (s=85\%)}

\addplot[
    blue,
    mark=triangle,
    mark size=2pt,
]
coordinates {
    (61926655861, 1226.04)
    (65837858399, 1247.46)
    (65837938724, 1338.64)
    (65837996100, 1435.19)
};
\addlegendentry{After GroupBy (s=15\%)}

\addplot[
    blue,
    mark=triangle*,
    mark size=2pt,
]
coordinates {
    (61926655861, 1791.05)
    (65837858399, 1825)
    (65837901431, 1867.9)
};
\addlegendentry{After GroupBy (s=85\%)}

\addplot[
    orange,
    mark=o,
    mark size=2pt,
]
coordinates {
    (1863088, 102.1)
    (3530060, 132.6)
    (3962734, 235.72)
    (3963770, 381.3)
};
\addlegendentry{After Join (s=15\%)}

\addplot[
    orange,
    mark=*,
    mark size=2pt,
]
coordinates {
    (1863088, 1543.71)
    (3530060, 1721.11)
    (3962061, 1904)
};
\addlegendentry{After Join (s=85\%)}

\addplot[red] coordinates {
    (1, 1753.5)
    (100000000000, 1753.5)
};
\addlegendentry{Fully Oblivious}

\addplot[
    red,
    mark=star,
    mark size=3pt,
    only marks
]
coordinates {
    (1, 2.25)
};
\addlegendentry{Fully Revealed}

\end{axis}
\end{tikzpicture}
\caption{Different Resizer placement rules for modified TPC-H Q3, with average SQL operator selectivity of s=15\%, respectively s=85\%, impacts RtR (depicted: minimum RtR across operators in the query). Greedily adding a \resizer~ after each oblivious operator isn't always optimal. The \resizer~integration method impacts both performance and privacy.}
\label{fig:privacy_vs_runtime_points}
\end{figure}

\begin{table*}[t]
\centering
\resizebox{\textwidth}{!}{
\begin{tabular}{|l|l|l|l|l|l|l|l|l|l|}
\hline
\textbf{System} & \textbf{Security} & \textbf{MPC Framework (Domain)} & \textbf{IRS Protection (Operators)} & Configurable Protection & \textbf{Select Algorithm} & \textbf{Join Algorithm} & \textbf{Group By Algorithm} \\
\hline
SMCQL~\cite{smcql} & Semi-Honest & Oblivm~\cite{liu2015oblivm_Oblivm} (Boolean) & Fully Oblivious (All) & \emptycirc[0.8ex] & Sequential & NLJ & Sort-based \\
\hline
Shrinkwrap~\cite{shrinkwrap_bater2018shrinkwrap} & Semi-Honest & EMP-toolkit~\cite{wang2016emp_emptoolkit} (Boolean) & DP-based Trimming (All) & \halfcircbottom[0.8ex] & Sequential & NLJ   & Sort-based  \\
\hline
Secretflow~\cite{scql_fangsecretflow} & Semi-Honest & SCQL~\cite{scql} (Boolean) & Fully Oblivious or Revealed (Join, Group by) & \halfcircbottom[0.8ex] & Sequential  & PSI-Join  & Sort-based  \\
\hline
Secrecy~\cite{secrecy_liagouris2023secrecy} & Semi-Honest & Self-implemented (Hybrid) & Fully Oblivious (All) & \emptycirc[0.8ex] & Sequential  & NLJ  & Sort-based  \\
\hline
\textsc{\mywrap} (this work) & Semi-Honest & MP-SPDZ~\cite{keller2020mp-spdz} (Hybrid) & Flexible Trimming (All) & \fullcirc[0.8ex] & Sequential  & NLJ  & Sort-based  \\
\hline

\end{tabular}
}
\caption{Comparison to related work. \textsc{\mywrap}~is the only system offering a flexible and efficient Resizer, enabling users to balance privacy and performance.}
\vspace{-2ex}
\label{tab:comparison}
\end{table*}

Key insights from this experiment can be summarized as follows: First, inserting \resizer~after every operator improves runtime but severely reduces privacy (low RtR), as fewer rounds are needed to infer the true intermediate result size of the weakest operator in the query plan. Therefore, a more targeted placement of \resizer~can yield comparable performance gains while achieving higher RtR. For example, the \textit{After Join} rule has runtime similar to the \textit{After all} rule, but with better RtR.

Second, placing \resizer~after GroupBy operators high up the query tree results in higher RtR but worse performance compared to placing it after Join operators. This suggests that placing \resizer~earlier in the query plan is perhaps a generally better choice.

Third, as the average selectivity of SQL operators increases, the effectiveness of \resizer~diminishes: execution time worsens while RtR remains similar. For example, the \textit{After GroupBy} and \textit{After Join} rules exhibit worse runtimes than the fully oblivious baseline under high selectivity.

Finally, the choice of distribution, i.e., the number of filler tuples added to the true intermediate result, clearly has a significant impact on performance. Adding more filler tuples has a super-linear increase in runtime. However, in most cases, it does not always lead to proportionally better RtR. For all plotted placement rules, beyond a certain point, retaining more filler tuples no longer meaningfully increases the number of rounds required to infer true intermediate result sizes.

\subsection{Analytical Queries from Related Work}
\label{subsec:overall_perf}

\textbf{HealthLnk Queries in \textsc{\mywrap}.} We investigate the runtime of HealthLnk queries \cite{PCORI2015_healthlnk_query} used in related work \cite{smcql,shrinkwrap_bater2018shrinkwrap,saqe_bater2020SAQE}, that are indicative of clinical data research methodologies~\cite{hernandez2015adaptable_healthlnk_shrinkwrap}. In \Cref{fig:healthlnk-queries} we present the execution times of the Comorbidity Study, Dosage Study, Aspirin Count, and 3-join queries. Unless otherwise specified, each table has been populated with $N=1000$ synthetic tuples, in a way that filter and join selectivities are $10\%N$ per operator.

~\Cref{fig:healthlnk-queries} compares the runtimes (in seconds) of \textsc{\mywrap} in the following settings: (i) fully oblivious execution without intermediate trimming (red diamonds), resulting in fully oblivious intermediate result sizes (IRS), (ii) \textsc{\mywrap} using \resizer~operators with a probability distribution that in expectation keeps $\eta=30\%N$ filler tuples (orange triangles), (iii) \textsc{\mywrap} with $\eta=10\%N$ (blue squares), and (iv) oblivious execution of operators but trimming all filler tuples (green circles), effectively revealing the IRS. We placed a \resizer~operator after each operator in the query, except for the last operator.

\begin{figure}[t]
  \centering
  \small
    \begin{tikzpicture}
        \pgfplotstableread{
        Query FullyRevealed Reflex Shrinkwrap FullyOblivious reflex_50 reflex_1 plaintext
        0   1.52   2.1   2.6   3.13   2.9   0   0
        1   0.42   2.8   8.3   588  19.4   0   0
        2   0.37   3.01   10.13  608.9   22   0   0
        3   1.2  9      29    2579163   65   6.7   0
        }\datatable
    
        \begin{axis}[
            clip=false,
            enlarge x limits=0.27,
            ymode=log,
            width=0.95\columnwidth, 
            height=5cm,
            ylabel={Execution Time [s]},
            xtick=data,
            xticklabels={Comorbidity, Dosage Study, Aspirin Count, 3-Join},
            grid=major,
            legend style={
                at={(.378,1)},
                anchor=north east,
                font=\tiny,
                legend cell align=left,
                row sep=0pt,
                scale=0.6,
            },
            legend image post style={
            scale=0.6,
            line width=0.15pt,
            mark size=1.5pt},
            log basis y=10,
            x tick label style={rotate=30,anchor=east},  
            reverse legend,
        ]
    
        \addplot[only marks, color=green, mark=o, mark size=3pt] 
        table[x=Query, y=FullyRevealed] {\datatable};
        \addlegendentry{Revealed IRS}
    
        \addplot[only marks, color=blue, mark=square, mark size=3pt] 
        table[x=Query, y=Reflex] {\datatable};
        \addlegendentry{$E(\eta)=10\%N$}
    
        \addplot[only marks, color=orange, mark=triangle*, mark size=3pt] 
        table[x=Query, y=Shrinkwrap] {\datatable};
        \addlegendentry{$E(\eta)=30\%N$}
    
        \addplot[only marks, color=red, mark=diamond*, mark size=3pt] 
        table[x=Query, y=FullyOblivious] {\datatable};
        \addlegendentry{Fully Oblivious IRS}
    
        \draw[->, color=red,shorten >=1mm,shorten <=1mm] (axis cs:2.6,3e5) node[right, align=left, font=\footnotesize] at (axis cs:2,2.5e5) {Estimated} -- (axis cs:3,2e6); 

        \pgfplotstablegetelem{0}{FullyRevealed}\of\datatable
        \let\revealed\pgfplotsretval

        \node[anchor=west, font=\footnotesize, color=blue] at (axis cs:0.1, \revealed*1.5) {\pgfmathprintnumber[precision=1] 1.38x};

        \node[anchor=west, font=\footnotesize, color=orange] at (axis cs:0.1, \revealed*3.5) {\pgfmathprintnumber[precision=1] 1.71x};

        \node[anchor=west, font=\footnotesize, color=red] at (axis cs:0.1, \revealed*8.5) {\pgfmathprintnumber[precision=1] 2.1x};

        \pgfplotstablegetelem{1}{FullyRevealed}\of\datatable
        \let\revealed\pgfplotsretval

        \node[anchor=west, font=\footnotesize, color=blue] at (axis cs:1.1, \revealed*2.8) {\pgfmathprintnumber[precision=1] 6.67x};

        \node[anchor=west, font=\footnotesize, color=orange] at (axis cs:1.1, \revealed*8) {\pgfmathprintnumber[precision=1] 19.8x};

        \node[anchor=west, font=\footnotesize, color=red] at (axis cs:1.1, \revealed*578) {\pgfmathprintnumber[precision=1] 1400x};

        \pgfplotstablegetelem{2}{FullyRevealed}\of\datatable
        \let\revealed\pgfplotsretval

        \node[anchor=west, font=\footnotesize, color=blue] at (axis cs:2.1, \revealed*2.6) {\pgfmathprintnumber[precision=1] 8.1x};
        
        \node[anchor=west, font=\footnotesize, color=orange] at (axis cs:2.1, \revealed*10.56) {\pgfmathprintnumber[precision=1] 27.4x};

        \node[anchor=west, font=\footnotesize, color=red] at (axis cs:2.1, \revealed*578) {\pgfmathprintnumber[precision=1] 1646x};

        \pgfplotstablegetelem{3}{FullyRevealed}\of\datatable
        \let\revealed\pgfplotsretval

        \node[anchor=west, font=\footnotesize, color=blue] at (axis cs:3.1, \revealed*7.5) {\pgfmathprintnumber[precision=1] 7.5x};

        \node[anchor=west, font=\footnotesize, color=orange] at (axis cs:3.1, \revealed*29) {\pgfmathprintnumber[precision=1] 24.2x};

    
    \end{axis}
    \end{tikzpicture}
    \vspace{-3ex}
    \caption{We apply \resizer~ to the HealthLnK queries (synthetic data with table sizes of 1000 and fixed selectivity of 10\% per operator) and show that, when trimming filler tuples, performance is dramatically increased. If no fillers are trimmed, query runtimes increase by more than 1000x.}
    \label{fig:healthlnk-queries}
    \vspace{-5ex}
\end{figure}

The results show that reducing IRS leads to significantly reduced runtime for queries with joins (Dosage Study, Aspirin Count, 3-Join). Since the Comorbidity query does not involve a join operation, it is less affected by ballooning data sizes, benefiting only modestly from trimming. Note that the runtime of the 3-Join query under a fully oblivious setting is estimated (it could not be executed on our platform due to memory limitations). We estimate the runtime by measuring its runtime with smaller table sizes and extrapolating using the trend shown in~\Cref{fig:3join_qp}.
Overall, depending on the complexity of the queries to be executed -- and in particular the number of joins -- \textsc{\mywrap} can substantially boost performance when trimming away fillers.

\textbf{Comparison to Related Work.} 
Table~\ref{tab:comparison} compares \textsc{\mywrap}~with representative semi-honest analytics systems that use MPC. The table shows i) choice of MPC Framework and computational domain, which has an important effect on overall performance, ii) a summary of how these frameworks protect intermediate result size and whether they expose some user-facing parameters for relaxing these protections, and iii) the underlying algorithms used for filter, join, and group by. Due to the numerous design differences across frameworks and variations in implementation, it is relatively challenging to provide an apples-to-apples comparison of these systems. The goal of this table is to provide an overview and show that \textsc{\mywrap}~is based on a comparable foundation to the state of the art. 
In terms of baseline performance of the MPC frameworks, given the analytics use-case, those that operate in an arithmetic or hybrid model are expected to be significantly faster than those using solely binary: SMCQL~\cite{smcql} and Shrinkwrap~\cite{shrinkwrap_bater2018shrinkwrap} rely on Boolean-circuit frameworks (Oblivm~\cite{liu2015oblivm_Oblivm} and EMP-tookit~\cite{wang2016emp_emptoolkit}). SecretFlow~\cite{scql_fangsecretflow} also uses a Boolean domain in SCQL~\cite{scql}. Secrecy and \textsc{\mywrap} execute in hybrid mode, combining arithmetic and Boolean domains for improved efficiency.

In terms of protecting the intermediate result size, SMCQL~\cite{smcql} and Secrecy~\cite{secrecy_liagouris2023secrecy} enforce full obliviousness, incurring a high cost on operations such as joins and group by. Shrinkwrap~\cite{shrinkwrap_bater2018shrinkwrap} introduces a differentially private (DP) mechanism to reduce overhead. The user can influence the trimming behavior indirectly through the privacy budget and other parameters of the DP mechanism.  Secretflow~\cite{scql_fangsecretflow} can operate in two modes: either keeping IRS oblivious, without performance benefits, or exposing it fully. As the table shows, Secretflow offers additional security relaxations that users can choose, allowing, for instance, parties to learn something about the contents of a join result. However, this results in semi-oblivious execution of core operators, which no other system compromises on.
\textsc{\mywrap} is the only system with flexible IRS protection based on user-defined distributions. These distributions could even be used to provide the DP guarantees from related work.

Query processing happens in a similar fashion in all systems: they employ sequential selection and sort-based group-by~\cite{graefe1993query_evaluation_sort_groupby}. For joins, most use a nested-loop join (NLJ), which iteratively compares each pair of tuples from the input relations, whereas SecretFlow~\cite{scql_fangsecretflow} adopts a PSI-join, i.e., a join based on private set intersection~\cite{huberman1999enhancing_ecdh_psi_join} that securely identifies matching join keys between two datasets -- this algorithm, while only usable for key-to-key joins, has the better asymptotic complexity than NLJ. Overall, \textsc{\mywrap}~uses algorithms similar to the state of the art and leverages the maturity and efficiency of the underlying MP-SPDZ framework.

\section{Summary}
We presented \textsc{\mywrap}, an efficient and flexible approach for trimming oblivious intermediate results in MPC-based query execution. 

At the core of \textsc{\mywrap}~is the \resizer~ that can be inserted after any oblivious operator to trim its intermediate result size, without requiring query rewrites or changes to upstream operators. The decision of how many filler tuples to trim is taken using user-defined probability distributions. 
These allow capturing different points in the trade-off space. To enable a statistically-grounded comparison between distributions, we introduce the Round to Recover (RtR) metric, which applies to any distribution and helps planners select the most secure strategy within a given time budget. 



We evaluated \textsc{\mywrap}~using the queries from related work~\cite{smcql,shrinkwrap_bater2018shrinkwrap,saqe_bater2020SAQE} and achieved runtime reduction thanks to trimmed intermediate results in similar orders of magnitude to related work. Generally, \textsc{\mywrap}~is even faster thanks to its efficient implementation in MP-SPDZ. 
\textsc{\mywrap}~paves the way for future query optimizers that consider both privacy and performance, as it allows future query optimizers to compile SQL queries with \resizer~included, and, from our results, cost models can be derived on the effect of filler tuples on operator runtime.

\section*{Artifacts}
\label{sec:artifacts}
The source code and data used in this paper are available at:\\ \url{https://github.com/DataManagementLab/reflex-smpc-analytics}


\bibliographystyle{ACM-Reference-Format}
\bibliography{ref}


\begin{thebibliography}{44}


\ifx \showCODEN    \undefined \def \showCODEN     #1{\unskip}     \fi
\ifx \showDOI      \undefined \def \showDOI       #1{#1}\fi
\ifx \showISBNx    \undefined \def \showISBNx     #1{\unskip}     \fi
\ifx \showISBNxiii \undefined \def \showISBNxiii  #1{\unskip}     \fi
\ifx \showISSN     \undefined \def \showISSN      #1{\unskip}     \fi
\ifx \showLCCN     \undefined \def \showLCCN      #1{\unskip}     \fi
\ifx \shownote     \undefined \def \shownote      #1{#1}          \fi
\ifx \showarticletitle \undefined \def \showarticletitle #1{#1}   \fi
\ifx \showURL      \undefined \def \showURL       {\relax}        \fi
\providecommand\bibfield[2]{#2}
\providecommand\bibinfo[2]{#2}
\providecommand\natexlab[1]{#1}
\providecommand\showeprint[2][]{arXiv:#2}

\bibitem[\protect\citeauthoryear{??}{scq}{2024}]%
        {scql}
 \bibinfo{year}{2024}\natexlab{}.
\newblock \showarticletitle{SecretFlow-SCQL: {A} Secure Collaborative Query pLatform.}
\newblock \bibinfo{journal}{\emph{Published in Proc. VLDB Endow.}} \bibinfo{volume}{17}, \bibinfo{number}{12} (\bibinfo{year}{2024}), \bibinfo{pages}{3987--4000}.
\newblock
\urldef\tempurl%
\url{https://github.com/secretflow/scql}
\showURL{%
\tempurl}


\bibitem[\protect\citeauthoryear{Araki, Furukawa, Lindell, Nof, and Ohara}{Araki et~al\mbox{.}}{2016}]%
        {RSS_replicated_secret_sharing_araki2016high}
\bibfield{author}{\bibinfo{person}{Toshinori Araki}, \bibinfo{person}{Jun Furukawa}, \bibinfo{person}{Yehuda Lindell}, \bibinfo{person}{Ariel Nof}, {and} \bibinfo{person}{Kazuma Ohara}.} \bibinfo{year}{2016}\natexlab{}.
\newblock \showarticletitle{High-throughput semi-honest secure three-party computation with an honest majority}. In \bibinfo{booktitle}{\emph{Proceedings of the 2016 ACM SIGSAC Conference on Computer and Communications Security}}. \bibinfo{pages}{805--817}.
\newblock


\bibitem[\protect\citeauthoryear{Asharov, Hamada, Ikarashi, Kikuchi, Nof, Pinkas, Takahashi, and Tomida}{Asharov et~al\mbox{.}}{2022}]%
        {Asharov22}
\bibfield{author}{\bibinfo{person}{Gilad Asharov}, \bibinfo{person}{Koki Hamada}, \bibinfo{person}{Dai Ikarashi}, \bibinfo{person}{Ryo Kikuchi}, \bibinfo{person}{Ariel Nof}, \bibinfo{person}{Benny Pinkas}, \bibinfo{person}{Katsumi Takahashi}, {and} \bibinfo{person}{Junichi Tomida}.} \bibinfo{year}{2022}\natexlab{}.
\newblock \showarticletitle{Efficient Secure Three-Party Sorting with Applications to Data Analysis and Heavy Hitters}. In \bibinfo{booktitle}{\emph{Proceedings of the 2022 ACM SIGSAC Conference on Computer and Communications Security}} (Los Angeles, CA, USA) \emph{(\bibinfo{series}{CCS '22})}. \bibinfo{publisher}{Association for Computing Machinery}, \bibinfo{address}{New York, NY, USA}, \bibinfo{pages}{125–138}.
\newblock
\showISBNx{9781450394505}
\urldef\tempurl%
\url{https://doi.org/10.1145/3548606.3560691}
\showDOI{\tempurl}


\bibitem[\protect\citeauthoryear{Bater, Elliott, Eggen, Goel, Kho, and Rogers}{Bater et~al\mbox{.}}{2016}]%
        {smcql}
\bibfield{author}{\bibinfo{person}{Johes Bater}, \bibinfo{person}{Gregory Elliott}, \bibinfo{person}{Craig Eggen}, \bibinfo{person}{Satyender Goel}, \bibinfo{person}{Abel Kho}, {and} \bibinfo{person}{Jennie Rogers}.} \bibinfo{year}{2016}\natexlab{}.
\newblock \bibinfo{title}{SMCQL: Secure Querying for Federated Databases}.
\newblock
\newblock


\bibitem[\protect\citeauthoryear{Bater, He, Ehrich, Machanavajjhala, and Rogers}{Bater et~al\mbox{.}}{2018}]%
        {shrinkwrap_bater2018shrinkwrap}
\bibfield{author}{\bibinfo{person}{Johes Bater}, \bibinfo{person}{Xi He}, \bibinfo{person}{William Ehrich}, \bibinfo{person}{Ashwin Machanavajjhala}, {and} \bibinfo{person}{Jennie Rogers}.} \bibinfo{year}{2018}\natexlab{}.
\newblock \showarticletitle{Shrinkwrap: efficient sql query processing in differentially private data federations}.
\newblock \bibinfo{journal}{\emph{Proceedings of the VLDB Endowment}} \bibinfo{volume}{12}, \bibinfo{number}{3} (\bibinfo{year}{2018}).
\newblock


\bibitem[\protect\citeauthoryear{Bater, Park, He, Wang, and Rogers}{Bater et~al\mbox{.}}{2020}]%
        {saqe_bater2020SAQE}
\bibfield{author}{\bibinfo{person}{Johes Bater}, \bibinfo{person}{Yongjoo Park}, \bibinfo{person}{Xi He}, \bibinfo{person}{Xiao Wang}, {and} \bibinfo{person}{Jennie Rogers}.} \bibinfo{year}{2020}\natexlab{}.
\newblock \showarticletitle{Saqe: practical privacy-preserving approximate query processing for data federations}.
\newblock \bibinfo{journal}{\emph{Proceedings of the VLDB Endowment}} \bibinfo{volume}{13}, \bibinfo{number}{12} (\bibinfo{year}{2020}), \bibinfo{pages}{2691--2705}.
\newblock


\bibitem[\protect\citeauthoryear{Bian, Zhang, Pan, Mao, Zhao, Jin, and Guan}{Bian et~al\mbox{.}}{2023}]%
        {2023HE3db}
\bibfield{author}{\bibinfo{person}{Song Bian}, \bibinfo{person}{Zhou Zhang}, \bibinfo{person}{Haowen Pan}, \bibinfo{person}{Ran Mao}, \bibinfo{person}{Zian Zhao}, \bibinfo{person}{Yier Jin}, {and} \bibinfo{person}{Zhenyu Guan}.} \bibinfo{year}{2023}\natexlab{}.
\newblock \showarticletitle{HE3DB: An Efficient and Elastic Encrypted Database Via Arithmetic-And-Logic Fully Homomorphic Encryption}. In \bibinfo{booktitle}{\emph{Proceedings of the 2023 ACM SIGSAC Conference on Computer and Communications Security}}. \bibinfo{pages}{2930--2944}.
\newblock


\bibitem[\protect\citeauthoryear{Brasser, M{\"u}ller, Dmitrienko, Kostiainen, Capkun, and Sadeghi}{Brasser et~al\mbox{.}}{2017}]%
        {sgx_sidechannel_brasser2017software}
\bibfield{author}{\bibinfo{person}{Ferdinand Brasser}, \bibinfo{person}{Urs M{\"u}ller}, \bibinfo{person}{Alexandra Dmitrienko}, \bibinfo{person}{Kari Kostiainen}, \bibinfo{person}{Srdjan Capkun}, {and} \bibinfo{person}{Ahmad-Reza Sadeghi}.} \bibinfo{year}{2017}\natexlab{}.
\newblock \showarticletitle{Software grand exposure:{SGX} cache attacks are practical}. In \bibinfo{booktitle}{\emph{11th USENIX workshop on offensive technologies (WOOT 17)}}.
\newblock


\bibitem[\protect\citeauthoryear{Chang, Xie, Wang, and Li}{Chang et~al\mbox{.}}{2022}]%
        {chang2022towards_oblivious_join}
\bibfield{author}{\bibinfo{person}{Zhao Chang}, \bibinfo{person}{Dong Xie}, \bibinfo{person}{Sheng Wang}, {and} \bibinfo{person}{Feifei Li}.} \bibinfo{year}{2022}\natexlab{}.
\newblock \showarticletitle{Towards practical oblivious join}. In \bibinfo{booktitle}{\emph{Proceedings of the 2022 International Conference on Management of Data}}. \bibinfo{pages}{803--817}.
\newblock


\bibitem[\protect\citeauthoryear{Commission}{Commission}{[n.d.]}]%
        {dpbd}
\bibfield{author}{\bibinfo{person}{European Commission}.} \bibinfo{year}{[n.d.]}\natexlab{}.
\newblock \bibinfo{howpublished}{\url{https://commission.europa.eu/law/law-topic/data-protection/rules-business-and-organisations/obligations/what-does-data-protection-design-and-default-mean_en}}.
\newblock


\bibitem[\protect\citeauthoryear{Costan and Devadas}{Costan and Devadas}{2016}]%
        {costan16sgx}
\bibfield{author}{\bibinfo{person}{Victor Costan} {and} \bibinfo{person}{Srinivas Devadas}.} \bibinfo{year}{2016}\natexlab{}.
\newblock \bibinfo{title}{Intel {{SGX Explained}}}.
\newblock
\newblock
\urldef\tempurl%
\url{https://eprint.iacr.org/2016/086.pdf}
\showURL{%
\tempurl}


\bibitem[\protect\citeauthoryear{De~Meulemeester, Oswald, Verbauwhede, and Van~Bulck}{De~Meulemeester et~al\mbox{.}}{2026}]%
        {batteringramsp26_sgx_attack_battery_ram}
\bibfield{author}{\bibinfo{person}{Jesse De~Meulemeester}, \bibinfo{person}{David Oswald}, \bibinfo{person}{Ingrid Verbauwhede}, {and} \bibinfo{person}{Jo Van~Bulck}.} \bibinfo{year}{2026}\natexlab{}.
\newblock \showarticletitle{{Battering RAM}: Low-Cost Interposer Attacks on Confidential Computing via Dynamic Memory Aliasing}. In \bibinfo{booktitle}{\emph{47th {IEEE} Symposium on Security and Privacy ({S\&P})}}.
\newblock


\bibitem[\protect\citeauthoryear{Dimitriou and Michalas}{Dimitriou and Michalas}{2014}]%
        {dimitriou2014multi}
\bibfield{author}{\bibinfo{person}{Tassos Dimitriou} {and} \bibinfo{person}{Antonis Michalas}.} \bibinfo{year}{2014}\natexlab{}.
\newblock \showarticletitle{Multi-party trust computation in decentralized environments in the presence of malicious adversaries}.
\newblock \bibinfo{journal}{\emph{Ad Hoc Networks}}  \bibinfo{volume}{15} (\bibinfo{year}{2014}), \bibinfo{pages}{53--66}.
\newblock


\bibitem[\protect\citeauthoryear{Escudero, Ghosh, Keller, Rachuri, and Scholl}{Escudero et~al\mbox{.}}{2020}]%
        {escudero2020improved_edabits}
\bibfield{author}{\bibinfo{person}{Daniel Escudero}, \bibinfo{person}{Satrajit Ghosh}, \bibinfo{person}{Marcel Keller}, \bibinfo{person}{Rahul Rachuri}, {and} \bibinfo{person}{Peter Scholl}.} \bibinfo{year}{2020}\natexlab{}.
\newblock \showarticletitle{Improved primitives for MPC over mixed arithmetic-binary circuits}. In \bibinfo{booktitle}{\emph{Annual international cryptology conference}}. Springer, \bibinfo{pages}{823--852}.
\newblock


\bibitem[\protect\citeauthoryear{Eskandarian and Boneh}{Eskandarian and Boneh}{2021}]%
        {linkage_attack_eskandarian2021clarion}
\bibfield{author}{\bibinfo{person}{Saba Eskandarian} {and} \bibinfo{person}{Dan Boneh}.} \bibinfo{year}{2021}\natexlab{}.
\newblock \showarticletitle{Clarion: Anonymous communication from multiparty shuffling protocols}.
\newblock \bibinfo{journal}{\emph{Cryptology ePrint Archive}} (\bibinfo{year}{2021}).
\newblock


\bibitem[\protect\citeauthoryear{Eskandarian and Zaharia}{Eskandarian and Zaharia}{2017}]%
        {sgx_oblidb_eskandarian2017oblidb}
\bibfield{author}{\bibinfo{person}{Saba Eskandarian} {and} \bibinfo{person}{Matei Zaharia}.} \bibinfo{year}{2017}\natexlab{}.
\newblock \showarticletitle{Oblidb: Oblivious query processing for secure databases}.
\newblock \bibinfo{journal}{\emph{arXiv preprint arXiv:1710.00458}} (\bibinfo{year}{2017}).
\newblock


\bibitem[\protect\citeauthoryear{Fang, Cao, Hua, Ma, Yu, Huang, Feng, Tan, Zan, Duan, et~al\mbox{.}}{Fang et~al\mbox{.}}{2024}]%
        {scql_fangsecretflow}
\bibfield{author}{\bibinfo{person}{Wenjing Fang}, \bibinfo{person}{Shunde Cao}, \bibinfo{person}{Guojin Hua}, \bibinfo{person}{Junming Ma}, \bibinfo{person}{Yongqiang Yu}, \bibinfo{person}{Qunshan Huang}, \bibinfo{person}{Jun Feng}, \bibinfo{person}{Jin Tan}, \bibinfo{person}{Xiaopeng Zan}, \bibinfo{person}{Pu Duan}, {et~al\mbox{.}}} \bibinfo{year}{2024}\natexlab{}.
\newblock \showarticletitle{SecretFlow-SCQL: A Secure Collaborative Query pLatform}.
\newblock  (\bibinfo{year}{2024}).
\newblock


\bibitem[\protect\citeauthoryear{Furukawa and Lindell}{Furukawa and Lindell}{2019}]%
        {Lindell19}
\bibfield{author}{\bibinfo{person}{Jun Furukawa} {and} \bibinfo{person}{Yehuda Lindell}.} \bibinfo{year}{2019}\natexlab{}.
\newblock \showarticletitle{Two-Thirds Honest-Majority MPC for Malicious Adversaries at Almost the Cost of Semi-Honest}. In \bibinfo{booktitle}{\emph{Proceedings of the 2019 ACM SIGSAC Conference on Computer and Communications Security}} \emph{(\bibinfo{series}{CCS '19})}. \bibinfo{publisher}{Association for Computing Machinery}.
\newblock
\showISBNx{9781450367479}


\bibitem[\protect\citeauthoryear{GDPR}{GDPR}{2016}]%
        {GDPR2016general}
\bibfield{author}{\bibinfo{person}{General Data Protection~Regulation GDPR}.} \bibinfo{year}{2016}\natexlab{}.
\newblock \showarticletitle{General data protection regulation}.
\newblock \bibinfo{journal}{\emph{Regulation (EU) 2016/679 of the European Parliament and of the Council of 27 April 2016 on the protection of natural persons with regard to the processing of personal data and on the free movement of such data, and repealing Directive 95/46/EC}} (\bibinfo{year}{2016}).
\newblock


\bibitem[\protect\citeauthoryear{Graefe}{Graefe}{1993}]%
        {graefe1993query_evaluation_sort_groupby}
\bibfield{author}{\bibinfo{person}{Goetz Graefe}.} \bibinfo{year}{1993}\natexlab{}.
\newblock \showarticletitle{Query evaluation techniques for large databases}.
\newblock \bibinfo{journal}{\emph{ACM Computing Surveys (CSUR)}} \bibinfo{volume}{25}, \bibinfo{number}{2} (\bibinfo{year}{1993}), \bibinfo{pages}{73--169}.
\newblock


\bibitem[\protect\citeauthoryear{Guimar{\~a}es, Borin, and Aranha}{Guimar{\~a}es et~al\mbox{.}}{2021}]%
        {FHE_guimaraes2021revisiting}
\bibfield{author}{\bibinfo{person}{Antonio Guimar{\~a}es}, \bibinfo{person}{Edson Borin}, {and} \bibinfo{person}{Diego~F Aranha}.} \bibinfo{year}{2021}\natexlab{}.
\newblock \showarticletitle{Revisiting the functional bootstrap in TFHE}.
\newblock \bibinfo{journal}{\emph{IACR Transactions on Cryptographic Hardware and Embedded Systems}} (\bibinfo{year}{2021}), \bibinfo{pages}{229--253}.
\newblock


\bibitem[\protect\citeauthoryear{Han, Zhang, Feng, Liu, and Li}{Han et~al\mbox{.}}{2022}]%
        {scape_han2022scape}
\bibfield{author}{\bibinfo{person}{Feng Han}, \bibinfo{person}{Lan Zhang}, \bibinfo{person}{Hanwen Feng}, \bibinfo{person}{Weiran Liu}, {and} \bibinfo{person}{Xiangyang Li}.} \bibinfo{year}{2022}\natexlab{}.
\newblock \showarticletitle{Scape: Scalable collaborative analytics system on private database with malicious security}. In \bibinfo{booktitle}{\emph{2022 IEEE 38th International Conference on Data Engineering (ICDE)}}. IEEE, \bibinfo{pages}{1740--1753}.
\newblock


\bibitem[\protect\citeauthoryear{He, Rogers, Bater, Machanavajjhala, Wang, and Wang}{He et~al\mbox{.}}{2021}]%
        {he2021practical_security_privacy_db}
\bibfield{author}{\bibinfo{person}{Xi He}, \bibinfo{person}{Jennie Rogers}, \bibinfo{person}{Johes Bater}, \bibinfo{person}{Ashwin Machanavajjhala}, \bibinfo{person}{Chenghong Wang}, {and} \bibinfo{person}{Xiao Wang}.} \bibinfo{year}{2021}\natexlab{}.
\newblock \showarticletitle{Practical security and privacy for database systems}. In \bibinfo{booktitle}{\emph{Proceedings of the 2021 International Conference on Management of Data}}. \bibinfo{pages}{2839--2845}.
\newblock


\bibitem[\protect\citeauthoryear{Hernandez, Fleurence, and Rothman}{Hernandez et~al\mbox{.}}{2015}]%
        {hernandez2015adaptable_healthlnk_shrinkwrap}
\bibfield{author}{\bibinfo{person}{Adrian~F Hernandez}, \bibinfo{person}{Rachael~L Fleurence}, {and} \bibinfo{person}{Russell~L Rothman}.} \bibinfo{year}{2015}\natexlab{}.
\newblock \bibinfo{title}{The ADAPTABLE Trial and PCORnet: shining light on a new research paradigm}.
\newblock , \bibinfo{numpages}{635--636}~pages.
\newblock


\bibitem[\protect\citeauthoryear{Huberman, Franklin, and Hogg}{Huberman et~al\mbox{.}}{1999}]%
        {huberman1999enhancing_ecdh_psi_join}
\bibfield{author}{\bibinfo{person}{Bernardo~A Huberman}, \bibinfo{person}{Matt Franklin}, {and} \bibinfo{person}{Tad Hogg}.} \bibinfo{year}{1999}\natexlab{}.
\newblock \showarticletitle{Enhancing privacy and trust in electronic communities}. In \bibinfo{booktitle}{\emph{Proceedings of the 1st ACM conference on Electronic commerce}}. \bibinfo{pages}{78--86}.
\newblock


\bibitem[\protect\citeauthoryear{Kaplan, Powell, and Woller}{Kaplan et~al\mbox{.}}{2016}]%
        {kaplan2016amd}
\bibfield{author}{\bibinfo{person}{David Kaplan}, \bibinfo{person}{Jeremy Powell}, {and} \bibinfo{person}{Tom Woller}.} \bibinfo{year}{2016}\natexlab{}.
\newblock \showarticletitle{AMD memory encryption}.
\newblock \bibinfo{journal}{\emph{White paper}}  \bibinfo{volume}{13} (\bibinfo{year}{2016}), \bibinfo{pages}{12}.
\newblock


\bibitem[\protect\citeauthoryear{Keller}{Keller}{2020}]%
        {keller2020mp-spdz}
\bibfield{author}{\bibinfo{person}{Marcel Keller}.} \bibinfo{year}{2020}\natexlab{}.
\newblock \showarticletitle{MP-SPDZ: A versatile framework for multi-party computation}. In \bibinfo{booktitle}{\emph{Proceedings of the 2020 ACM SIGSAC conference on computer and communications security}}. \bibinfo{pages}{1575--1590}.
\newblock


\bibitem[\protect\citeauthoryear{Keller and Scholl}{Keller and Scholl}{2014a}]%
        {waskman_efficient_secure_shuffle}
\bibfield{author}{\bibinfo{person}{Marcel Keller} {and} \bibinfo{person}{Peter Scholl}.} \bibinfo{year}{2014}\natexlab{a}.
\newblock \showarticletitle{Efficient, oblivious data structures for MPC}. In \bibinfo{booktitle}{\emph{Advances in Cryptology--ASIACRYPT 2014: 20th International Conference on the Theory and Application of Cryptology and Information Security, Kaoshiung, Taiwan, ROC, December 7-11, 2014, Proceedings, Part II 20}}. Springer, \bibinfo{pages}{506--525}.
\newblock


\bibitem[\protect\citeauthoryear{Keller and Scholl}{Keller and Scholl}{2014b}]%
        {keller2014efficient_mpspdz_shuffle}
\bibfield{author}{\bibinfo{person}{Marcel Keller} {and} \bibinfo{person}{Peter Scholl}.} \bibinfo{year}{2014}\natexlab{b}.
\newblock \showarticletitle{Efficient, oblivious data structures for MPC}. In \bibinfo{booktitle}{\emph{International Conference on the Theory and Application of Cryptology and Information Security}}. Springer, \bibinfo{pages}{506--525}.
\newblock


\bibitem[\protect\citeauthoryear{Lee, Shih, Gera, Kim, Kim, and Peinado}{Lee et~al\mbox{.}}{2017}]%
        {sgx_sidechannel_lee2017inferring}
\bibfield{author}{\bibinfo{person}{Sangho Lee}, \bibinfo{person}{Ming-Wei Shih}, \bibinfo{person}{Prasun Gera}, \bibinfo{person}{Taesoo Kim}, \bibinfo{person}{Hyesoon Kim}, {and} \bibinfo{person}{Marcus Peinado}.} \bibinfo{year}{2017}\natexlab{}.
\newblock \showarticletitle{Inferring fine-grained control flow inside {SGX} enclaves with branch shadowing}. In \bibinfo{booktitle}{\emph{26th USENIX Security Symposium (USENIX Security 17)}}. \bibinfo{pages}{557--574}.
\newblock


\bibitem[\protect\citeauthoryear{Liagouris, Kalavri, Faisal, and Varia}{Liagouris et~al\mbox{.}}{2023}]%
        {secrecy_liagouris2023secrecy}
\bibfield{author}{\bibinfo{person}{John Liagouris}, \bibinfo{person}{Vasiliki Kalavri}, \bibinfo{person}{Muhammad Faisal}, {and} \bibinfo{person}{Mayank Varia}.} \bibinfo{year}{2023}\natexlab{}.
\newblock \showarticletitle{{SECRECY}: Secure collaborative analytics in untrusted clouds}. In \bibinfo{booktitle}{\emph{20th USENIX Symposium on Networked Systems Design and Implementation (NSDI 23)}}. \bibinfo{pages}{1031--1056}.
\newblock


\bibitem[\protect\citeauthoryear{Liu, Wang, Nayak, Huang, and Shi}{Liu et~al\mbox{.}}{2015}]%
        {liu2015oblivm_Oblivm}
\bibfield{author}{\bibinfo{person}{Chang Liu}, \bibinfo{person}{Xiao~Shaun Wang}, \bibinfo{person}{Kartik Nayak}, \bibinfo{person}{Yan Huang}, {and} \bibinfo{person}{Elaine Shi}.} \bibinfo{year}{2015}\natexlab{}.
\newblock \showarticletitle{Oblivm: A programming framework for secure computation}. In \bibinfo{booktitle}{\emph{2015 IEEE Symposium on Security and Privacy}}. IEEE, \bibinfo{pages}{359--376}.
\newblock


\bibitem[\protect\citeauthoryear{Lutsch, El{-}Hindi, Heinrich, Ritter, Istv{\'{a}}n, and Binnig}{Lutsch et~al\mbox{.}}{2025}]%
        {lutsch_sgxbenchmark_edbt25}
\bibfield{author}{\bibinfo{person}{Adrian Lutsch}, \bibinfo{person}{Muhammad El{-}Hindi}, \bibinfo{person}{Matthias Heinrich}, \bibinfo{person}{Daniel Ritter}, \bibinfo{person}{Zsolt Istv{\'{a}}n}, {and} \bibinfo{person}{Carsten Binnig}.} \bibinfo{year}{2025}\natexlab{}.
\newblock \showarticletitle{Benchmarking Analytical Query Processing in Intel SGXv2}. In \bibinfo{booktitle}{\emph{Proceedings 28th International Conference on Extending Database Technology, {EDBT} 2025, Barcelona, Spain, March 25-28, 2025}}, \bibfield{editor}{\bibinfo{person}{Alkis Simitsis}, \bibinfo{person}{Bettina Kemme}, \bibinfo{person}{Anna Queralt}, \bibinfo{person}{Oscar Romero}, {and} \bibinfo{person}{Petar Jovanovic}} (Eds.). \bibinfo{publisher}{OpenProceedings.org}, \bibinfo{pages}{516--528}.
\newblock
\urldef\tempurl%
\url{https://doi.org/10.48786/EDBT.2025.41}
\showDOI{\tempurl}


\bibitem[\protect\citeauthoryear{McKeen, Alexandrovich, Berenzon, Rozas, Shafi, Shanbhogue, and Savagaonkar}{McKeen et~al\mbox{.}}{2013}]%
        {mckeen13innovative}
\bibfield{author}{\bibinfo{person}{Frank McKeen}, \bibinfo{person}{Ilya Alexandrovich}, \bibinfo{person}{Alex Berenzon}, \bibinfo{person}{Carlos~V. Rozas}, \bibinfo{person}{Hisham Shafi}, \bibinfo{person}{Vedvyas Shanbhogue}, {and} \bibinfo{person}{Uday~R. Savagaonkar}.} \bibinfo{year}{2013}\natexlab{}.
\newblock \showarticletitle{Innovative Instructions and Software Model for Isolated Execution}. In \bibinfo{booktitle}{\emph{Proceedings of the 2nd {{International Workshop}} on {{Hardware}} and {{Architectural Support}} for {{Security}} and {{Privacy}}}} \emph{(\bibinfo{series}{{{HASP}} '13})}. \bibinfo{publisher}{Association for Computing Machinery}, \bibinfo{address}{New York, NY, USA}, \bibinfo{pages}{1}.
\newblock
\showISBNx{978-1-4503-2118-1}
\urldef\tempurl%
\url{https://doi.org/10.1145/2487726.2488368}
\showDOI{\tempurl}


\bibitem[\protect\citeauthoryear{{Patient-Centered Outcomes Research Institute (PCORI)}}{{Patient-Centered Outcomes Research Institute (PCORI)}}{2015}]%
        {PCORI2015_healthlnk_query}
\bibfield{author}{\bibinfo{person}{{Patient-Centered Outcomes Research Institute (PCORI)}}.} \bibinfo{year}{2015}\natexlab{}.
\newblock \bibinfo{title}{Exchanging de-identified data between hospitals for city-wide health analysis in the Chicago Area HealthLNK data repository (HDR)}.
\newblock
\newblock


\bibitem[\protect\citeauthoryear{Priebe, Vaswani, and Costa}{Priebe et~al\mbox{.}}{2018}]%
        {enclavedb_sgx_priebe2018enclavedb}
\bibfield{author}{\bibinfo{person}{Christian Priebe}, \bibinfo{person}{Kapil Vaswani}, {and} \bibinfo{person}{Manuel Costa}.} \bibinfo{year}{2018}\natexlab{}.
\newblock \showarticletitle{EnclaveDB: A secure database using SGX}. In \bibinfo{booktitle}{\emph{2018 IEEE Symposium on Security and Privacy (SP)}}. IEEE, \bibinfo{pages}{264--278}.
\newblock


\bibitem[\protect\citeauthoryear{Roy~Chowdhury, Wang, He, Machanavajjhala, and Jha}{Roy~Chowdhury et~al\mbox{.}}{2020}]%
        {roy2020crypt_crypt_system_dp}
\bibfield{author}{\bibinfo{person}{Amrita Roy~Chowdhury}, \bibinfo{person}{Chenghong Wang}, \bibinfo{person}{Xi He}, \bibinfo{person}{Ashwin Machanavajjhala}, {and} \bibinfo{person}{Somesh Jha}.} \bibinfo{year}{2020}\natexlab{}.
\newblock \showarticletitle{Crypt$\epsilon$: Crypto-assisted differential privacy on untrusted servers}. In \bibinfo{booktitle}{\emph{Proceedings of the 2020 ACM SIGMOD International Conference on Management of Data}}. \bibinfo{pages}{603--619}.
\newblock


\bibitem[\protect\citeauthoryear{Sasy, Johnson, and Goldberg}{Sasy et~al\mbox{.}}{2023}]%
        {sasy2023waks_secure_shuffling_sorting}
\bibfield{author}{\bibinfo{person}{Sajin Sasy}, \bibinfo{person}{Aaron Johnson}, {and} \bibinfo{person}{Ian Goldberg}.} \bibinfo{year}{2023}\natexlab{}.
\newblock \showarticletitle{Waks-on/waks-off: Fast oblivious offline/online shuffling and sorting with waksman networks}. In \bibinfo{booktitle}{\emph{Proceedings of the 2023 ACM SIGSAC Conference on Computer and Communications Security}}. \bibinfo{pages}{3328--3342}.
\newblock


\bibitem[\protect\citeauthoryear{Seto, Duran, Amer, Chuang, van Schaik, Genkin, and Garman}{Seto et~al\mbox{.}}{2025}]%
        {wiretap_sgx_attack}
\bibfield{author}{\bibinfo{person}{Alex Seto}, \bibinfo{person}{Oytun~Kuday Duran}, \bibinfo{person}{Samy Amer}, \bibinfo{person}{Jalen Chuang}, \bibinfo{person}{Stephan van Schaik}, \bibinfo{person}{Daniel Genkin}, {and} \bibinfo{person}{Christina Garman}.} \bibinfo{year}{2025}\natexlab{}.
\newblock \showarticletitle{WireTap: Breaking Server SGX via DRAM Bus Interposition}. In \bibinfo{booktitle}{\emph{2025 SIGSAC Conference on Computer and Communications Security (CCS '25)}}. \bibinfo{publisher}{Association for Computing Machinery}.
\newblock
\urldef\tempurl%
\url{https://wiretap.fail}
\showURL{%
\tempurl}


\bibitem[\protect\citeauthoryear{Sohn, Jiang, Hammer, and Rogers}{Sohn et~al\mbox{.}}{2025}]%
        {sohn2025_alchemy}
\bibfield{author}{\bibinfo{person}{Donghyun Sohn}, \bibinfo{person}{Kelly Jiang}, \bibinfo{person}{Nicolas Hammer}, {and} \bibinfo{person}{Jennie Rogers}.} \bibinfo{year}{2025}\natexlab{}.
\newblock \showarticletitle{Alchemy: A Query Optimization Framework for Oblivious SQL}.
\newblock \bibinfo{journal}{\emph{Proceedings of the VLDB Endowment}} \bibinfo{volume}{18}, \bibinfo{number}{9} (\bibinfo{year}{2025}), \bibinfo{pages}{3021--3034}.
\newblock


\bibitem[\protect\citeauthoryear{Trieflinger}{Trieflinger}{2020}]%
        {Bosch_use_case}
\bibfield{author}{\bibinfo{person}{Sven Trieflinger}.} \bibinfo{year}{2020}\natexlab{}.
\newblock \bibinfo{booktitle}{\emph{Trustworthy computing – data sovereignty while connected}}.
\newblock
\urldef\tempurl%
\url{https://www.bosch.com/research/research-fields/digitalization-and-connectivity/research-on-security-and-privacy/trustworthy-computing-data-sovereignty-while-connected/}
\showURL{%
\tempurl}


\bibitem[\protect\citeauthoryear{Volgushev, Schwarzkopf, Getchell, Varia, Lapets, and Bestavros}{Volgushev et~al\mbox{.}}{2019}]%
        {conclave_volgushev2019conclave}
\bibfield{author}{\bibinfo{person}{Nikolaj Volgushev}, \bibinfo{person}{Malte Schwarzkopf}, \bibinfo{person}{Ben Getchell}, \bibinfo{person}{Mayank Varia}, \bibinfo{person}{Andrei Lapets}, {and} \bibinfo{person}{Azer Bestavros}.} \bibinfo{year}{2019}\natexlab{}.
\newblock \showarticletitle{Conclave: secure multi-party computation on big data}. In \bibinfo{booktitle}{\emph{Proceedings of the Fourteenth EuroSys Conference 2019}}. \bibinfo{pages}{1--18}.
\newblock


\bibitem[\protect\citeauthoryear{Wang, Malozemoff, and Katz}{Wang et~al\mbox{.}}{2016}]%
        {wang2016emp_emptoolkit}
\bibfield{author}{\bibinfo{person}{Xiao Wang}, \bibinfo{person}{Alex~J Malozemoff}, {and} \bibinfo{person}{Jonathan Katz}.} \bibinfo{year}{2016}\natexlab{}.
\newblock \bibinfo{title}{EMP-toolkit: Efficient MultiParty computation toolkit}.
\newblock
\newblock


\bibitem[\protect\citeauthoryear{Yanai, Deng, and Hellerstein}{Yanai et~al\mbox{.}}{2021}]%
        {senate_yanai2021senate}
\bibfield{author}{\bibinfo{person}{Rishabh Poddar Sukrit Kalra~Avishay Yanai}, \bibinfo{person}{Ryan Deng}, {and} \bibinfo{person}{Raluca Ada Popa Joseph~M Hellerstein}.} \bibinfo{year}{2021}\natexlab{}.
\newblock \showarticletitle{Senate: A maliciously-secure MPC platform for collaborative analytics}. In \bibinfo{booktitle}{\emph{30th USENIX Security Symposium (USENIX Security 21), Vancouver, BC}}.
\newblock


\end{thebibliography}


\end{document}